\documentclass[prd,showkeys,floatfix,twocolumn,amsmath,amssymb,floatfix]{revtex4}
\usepackage{amsmath}
\usepackage{graphicx}
\usepackage{float}
\usepackage{epstopdf}
\usepackage{multirow}
\usepackage{subfigure}
\usepackage{color}
\usepackage[colorlinks,citecolor=blue,anchorcolor=red,menucolor=red,linkcolor=red,filecolor=red,runcolor=red,urlcolor=blue,frenchlinks=red]{hyperref}
\usepackage{orcidlink}

\allowdisplaybreaks[3]

\begin{document}
%=====================================================================================
%=====================================================================================
\title{Triply heavy tetraquark states with different flavors}
%=====================================================================================
%=====================================================================================
%

\author{Hui-Min Yang\orcidlink{0000-0001-9673-5623}$^1$}
\email{hmyang@pku.edu.cn}
\author{Yao Ma\orcidlink{0000-0002-5868-1166}$^1$}
\email{yaoma@pku.edu.cn}
\author{Wei-Lin Wu\orcidlink{0009-0009-3480-8810}$^2$}
\email{wlwu@pku.edu.cn}
\author{Shi-Lin Zhu\orcidlink{0000-0002-4055-6906}$^1$}
\email{zhusl@pku.edu.cn}

\affiliation{
$^1$School of Physics and Center of High Energy Physics, Peking University, Beijing 100871, China \\
$^2$School of Physics, Peking University, Beijing 100871, China
}

\begin{abstract}
We investigate the $S$-wave triply heavy tetraquark systems, including the $QQ^{(\prime)}\bar Q^{(\prime)}\bar q$ and $QQ^{(\prime)}\bar Q^{(\prime)}\bar s$ configurations ($Q^{(\prime)} = c, b$, $q = u, d$) with spin-parity $J^P = 0^+$, $1^+$, and $2^+$, within the framework of the constituent quark model. We construct the wave functions for these systems, incorporating complete color-spin wave functions and spatial wave functions with three different Jacobi coordinate configurations. We use complex scaling and Gaussian expansion method to solve the complex-scaled four-body Schr\"{o}dinger equation and obtain possible exotic states. To analyze the spatial properties of the tetraquark states, we compute the root-mean-square (rms) radii, which help distinguish between meson-molecular and compact tetraquark states. We find that there do not exist any bound states in the $QQ^{(\prime)}\bar Q^{(\prime)}\bar q$ and $QQ^{(\prime)}\bar Q^{(\prime)}\bar s$ systems. However, we identify a possible molecular resonant state $T_{3c,2}(5819)$ in the $cc\bar c\bar q$ system with spin-parity $J^P = 2^+$, which lies slightly above the  $J/\psi D^*(2S)$ threshold and has a large rms radius around 2.2 fm. Furthermore, we obtain a series of compact resonant states, some of which exhibit spatial structure similar to the tritium atom in QED, where the light quark circles around the cluster of three heavy quarks.
The lowest resonant state in the triply heavy systems has a mass of $5634$ MeV, a width of $16$ MeV and spin-parity $J^P=1^+$. We suggest searching for this state in the $J/\psi D$, $\eta_c D^*$, and $J/\psi D^*$ decay channels.
\end{abstract}
\pacs{14.20.Mr, 12.38.Lg, 12.39.Hg}
\keywords{triply heavy tetraquark systems, constituent quark model, complex scaling method, Gaussian expansion method}
\maketitle
\pagenumbering{arabic}
%
%
%
%=====================================================================================
%=====================================================================================
\section{Introduction}\label{sec:intro}
%=====================================================================================
%=====================================================================================
%
Quantum Chromodynamics (QCD) predicts the existence of color singlets, such as conventional mesons and baryons in the traditional quark model, multiquark states, glueballs, and hybrids~\cite{Rapidis:1977cv,Belle:2003nnu,BESIII:2023wfi,E852:2004gpn,CLEO:2011upl,Qi:2025hwd}. The exploration of exotic hadron states beyond conventional hadrons has become one of the most intriguing and exciting research areas in hadron physics, as it offers valuable insights into non-perturbative QCD. The concept of multiquark states was first proposed during the early development of the quark model~\cite{Gell-Mann:1964ewy,Zweig:1964ruk}. However, experimental progress in this area remained very limited for a long time. A significant breakthrough occurred in 2003 when the Belle Collaboration reported the discovery of a new charmonium-like state, $X(3872)$, observed in the exclusive $B^{\pm} \to K^{\pm} \pi^{+} \pi^{-} J/\psi$ decays~\cite{Belle:2003nnu}. Following this pivotal discovery, a growing number of $XYZ$ states have been experimentally identified in recent years. Notably, many of these newly discovered states contain at least one heavy quark or antiquark. Based on the number of heavy quark components, these states can be classified into three distinct categories: those containing one, two, or four heavy quarks (antiquarks). We provide a review of each of these categories.

\begin{itemize}
\item  Systems containing a singly heavy quark (antiquark), such as the charmed-strange states $D_{s0}(2317)$ and $D_{s1}(2460)$, have attracted significant attention. The $D_{s0}(2317)$ state was first observed by the BaBar collaboration in the $D_s^+\pi^0$ invariant mass spectrum in 2003~\cite{BaBar:2003oey}, and was later confirmed by several other collaborations, including CLEO and Belle~\cite{CLEO:2003ggt,Belle:2003guh}. Additionally, the CLEO collaboration identified another narrow peak, $D_{s1}(2460)$, in the $D^{+}_s\pi^0$ invariant mass spectrum~\cite{CLEO:2003ggt}. Several theoretical works have attempted to explain the $D_{s0}(2317)$ and $D_{s1}(2460)$ as $P$-wave charmed-strange mesons~\cite{Godfrey:2003kg,Liu:2020ruo}. However, their measured masses are significantly lower than those predicted by the traditional quark model~\cite{Godfrey:1985xj}, which has led to increased interest in investigating their internal structures. Various theoretical interpretations have been proposed, including the possibility that these states are molecular configurations, such as $D K$ or $D^* K$~\cite{Zhang:2006ix,Xiao:2016hoa,Guo:2006rp,Navarra:2015iea,Barnes:2003dj,Kolomeitsev:2003ac,Hofmann:2003je,Rosner:2006vc,Guo:2006fu} or compact tetraquark states~\cite{Wang:2006uba,Nielsen:2005ia,Tang:2016pcf,Terasaki:2005kc,Cheng:2003kg,Kim:2005gt,Chen:2004dy}.

\item Systems containing two heavy quarks (antiquarks), such as the charmonium-like state $X(3872)$ and doubly charmed tetraquark state $T_{cc}(3875)$, have become one of the most important and intriguing topics in exotic states. The $X(3872)$ state was first observed in 2003 by the Belle Collaboration in the $\pi^+\pi^- J/\psi$ invariant mass distributions from the exclusive decay process $B^\pm \to K^\pm \pi^+ \pi^- J/\psi$~\cite{Belle:2003nnu}, and subsequently confirmed by several other collaborations~\cite{CDF:2003cab,CDF:2005cfq,CDF:2006ocq,CDF:2009nxk,D0:2004zmu,CMS:2020eiw,LHCb:2011zzp,LHCb:2013kgk,LHCb:2017ymo,LHCb:2020xds,BESIII:2013fnz,BESIII:2019qvy,BESIII:2020nbj}. A considerable amount of theoretical work has interpreted it as $\chi_{c1}(2P)$ charmonium state~\cite{Meng:2007cx,Liu:2007uj,Wang:2010ej}. However, the fact that the $X(3872)$ is located very close to the $D\bar{D}^*$ threshold, leads to its consideration as a bound $D^{(*)}\bar{D}^{(*)}$ molecular state~\cite{Li:2012cs,Fleming:2007rp,Mehen:2015efa,Wang:2022qxe,Meng:2021kmi} or $D_s^{(*)}\bar{D}_s^{(*)}$~\cite{Meng:2020cbk,Barnes:1964pd}. A recent article points out that the $X(3872)$ originates either from the $D\bar D^*$ pole in a weak-coupling mode or from a shadow pole associated with the $\chi_{c1}(2P)$ in a strong-coupling mode~\cite{Wang:2024ytk}. The first doubly charmed tetraquark state, $T_{cc}(3875)^+$, was discovered by the LHCb Collaboration in the $D^0 D^0 \pi^+$ invariant mass spectrum~\cite{LHCb:2021vvq,LHCb:2021auc}. This state is notably narrow, with a mass extremely close to the $D^{*+} D^0$ threshold, exhibiting a binding energy of approximately $-300$~keV. The study of doubly heavy tetraquark systems has attracted significant attention from theorists, with various theoretical methods and models applied, including quark models~\cite{Meng:2024yhu,Meng:2023jqk,Wu:2024zbx,Ma:2023int}, lattice QCD~\cite{Lyu:2023xro,Whyte:2024ihh}, and recent progress summarized in several reviews~\cite{Dong:2021bvy,Wang:2023ovj,Feijoo:2021ppq,Huang:2021urd}. Other doubly heavy tetraquark states, such as $Z_c(3900)$~\cite{BESIII:2013ris,Belle:2013yex}, $Z_c(4020)$~\cite{BESIII:2013ouc}, $X(4050)$~\cite{Belle:2008qeq}, $X(4055)$~\cite{Belle:2014wyt}, $X(4100)$~\cite{LHCb:2018oeg}, $Z_c(4200)$~\cite{Belle:2014nuw}, $Z_c(4430)$~\cite{Belle:2007hrb}, $Z_b(10610)$~\cite{Belle:2011aa} and $Z_b(10650)$~\cite{Belle:2011aa}, etc, are considered as potential candidates for multiquark states or hadronic molecules~\cite{Yan:2021tcp}.

\item Systems containing four heavy quarks (antiquarks), such as the fully charmed tetraquark states $X(6900)$, $X(6600)$ and $X(7200)$, have made substantial strides in both experimental research and theoretical exploration.  In 2020, the first fully heavy tetraquark state, $X(6900)$, was observed in the double $J/\psi$ invariant mass distributions~\cite{LHCb:2020bwg} and subsequently confirmed by the CMS and ATLAS Collaborations~\cite{CMS:2023owd,ATLAS:2023bft}. Furthermore, several other resonant states, including $X(6600)$ and $X(7200)$, were also reported by CMS~\cite{CMS:2023owd} and ATLAS~\cite{ATLAS:2023bft} collaborations in the double $J/\psi$ invariant mass distributions. These recent experimental advancements have catalyzed extensive theoretical investigations into the nature of these states. Various interpretations have been explored, including compact tetraquark state~\cite{Wu:2024euj,Wu:2024hrv,liu:2020eha,Lu:2020cns,Tang:2024zvf} and gluonic tetraquark state~\cite{Wan:2020fsk,Wan:2024pet,Tang:2024kmh}, which have sparked further interest in multi-body systems composed of fully heavy quarks, driving further research into the dynamics and properties of such exotic states.

\end{itemize}

So far, no experimental reports have been made on triply heavy tetraquark states. However, substantial progress has been made in theoretical investigations in recent years. Some theoretical methods and models have been employed to explore the properties and dynamics of triply heavy tetraquark states, including  MIT bag model~\cite{Zhu:2023lbx}, effective field theory~\cite{Liu:2019mxw}, extended relativized quark model~\cite{Lu:2021kut}, extended chromomagnetic model~\cite{Weng:2021ngd}, Regge trajectories~\cite{Song:2024bkj}, and QCD sum rules~\cite{Jiang:2017tdc,Zhang:2024jvv}, etc. Ref.~\cite{Lu:2021kut} employed the extended relativized quark model to study all triply heavy tetraquarks above the corresponding meson-meson thresholds, finding no stable bound states. Ref.~\cite{Yang:2024nyc} identified triply charmed and bottom tetraquark resonances in the regions of $5.6-5.9$ GeV and $15.3-15.7$ GeV, respectively. In contrast, Ref.~\cite{Jiang:2017tdc} proposed the existence of bound states below the $\eta_b$ and $B$ ($B^*$) thresholds with quantum numbers $0^+$ ($1^+$).

In light of these discussions, we revisit and review these systems. In this work, we conduct a systematic study of $S$-wave triply heavy tetraquarks, including the $QQ^{(\prime)} \bar Q^{(\prime)}\bar q$ and $QQ^{(\prime)} \bar Q^{(\prime)}\bar s$ systems ($Q^{{\prime}}=c,b$, $q=u,d$) with spin-parity $J^P=0^+$, $1^+$ and $2^+$, within the constituent quark model. We construct the wave functions of these $S$-wave triply heavy tetraquark systems, incorporating a complete set of color-spin wave functions and spatial wave functions under three different Jacobi coordinate configurations. To predict possible exotic states, we use the complex scaling method~\cite{Aguilar:1971ve,Balslev:1971vb,Aoyama:2006hrz} (CSM) and Gaussian expansion method~\cite{Hiyama:2003cu} (GEM) to solve the complex-scaled four-body Schr\"{o}dinger equation, a technique that has proven effective in studies of tetraquark bound states~\cite{Meng:2023jqk,Chen:2023syh} and resonant states~\cite{Ma:2024vsi,Wu:2024euj,Wu:2024hrv,Wu:2024zbx}. To investigate the spatial properties of the tetraquark states, we compute the root-mean-square radii, which allow us to distinguish between meson-molecular states and compact tetraquark states based on their internal structures.

This paper is organized as follows. In Sec.~\ref{sec:formulation} we briefly introduce our theoretical framework, including the quark potential model, GEM and CSM. We construct the wave functions of $S$-wave triply heavy tetraquark systems. Additionally, we use the decomposed non-antisymmetric wave function to calculate the root-mean-square radii in systems of identical particles, which allows us to distinguish between meson-molecular and compact configurations. In Sec.~\ref{sec:results} we apply GEM and CSM to calculate the complex eigenenergies of $S$-wave triply heavy tetraquark systems and discuss the properties of these states. In Sec.~\ref{sec:summary} we conclude this paper.

%
%=====================================================================================
%=====================================================================================
\section{Formulation}
\label{sec:formulation}
%=====================================================================================
%=====================================================================================
%
\subsection{Hamiltonian}
The nonrelativisitic Hamiltonian of a triply heavy tetraquark system reads

\begin{eqnarray}
H&=& \sum_{i=1}^4({p_i^2\over 2m_i}+m_i)-T_{CM}+\sum_{i<j=1}^4 V_{ij}\, ,
\end{eqnarray}
where $m_i$ and $p_i$ represent the mass and momentum of the $i$-th quark, respectively. $T_{CM}$ denotes the center-of-mass kinematic energy, which has been subtracted in our calculation. $V_{ij}$ represents the interaction between the $i$-th and $j$-th quarks. In this work, we apply the AP1 quark potential model~\cite{Semay:1994ht,Silvestre-Brac:1996myf} to study triply heavy tetraquarks with different flavors. The AP1 potential includes both the one-gluon exchange (OGE) and a phenomenological $2/3$ power confinement interaction, as follows:
 
\begin{eqnarray}
\nonumber V_{ij}&=& V_{\rm {coul}}+V_{\rm {conf}}+V_{\rm{hyp}}+V_{\rm{cons}}
\\ \nonumber &=&-{3\over16} \lambda_i\cdot \lambda_j \Big(-{\kappa\over r_{ij}}+\lambda r^{2/3}_{ij}-\Lambda \\
&&+{8\pi \kappa^{\prime}\over 3 m_i m_j}{e^{-r_{ij}^2/r_0^2}\over \pi^{3/2}r_0^3}S_i\cdot S_j\Big)\, ,
\end{eqnarray}
The color matrix $\lambda$ is replaced by $-\lambda^*$ for an antiquark. The operator $S_i$ denotes the spin of the $i$-th quark. $r_{ij}$ refers to the relative position between the $i$-th and $j$-th quarks. The quantity $r_0$, defined as $r_0=A({2m_i m_j\over m_i+m_j})^{-B}$, is related to the reduced mass of two quarks $i$ and $j$. The potentials $V_{\rm{coul}}$, $V_{\rm {conf}}$, and $V_{\rm{hyp}}$ represent the OGE color Coulomb, the confinement, and hyperfine interactions, respectively. The OGE interaction leads to a contact hyperfine effect, which can be regularized using a Gaussian smearing function. The model parameters are presented in Table~\ref{tab:parameters} and taken from Ref.~\cite{Silvestre-Brac:1996myf}. The mass spectra and root-mean-square radii of heavy mesons calculated from AP1 potential are shown in Table~\ref{tab:mass}.
\begin{table*}[hbtp]
\begin{center}
\renewcommand{\arraystretch}{1.5}
\caption{The parameters in the AP1 quark model taken from Ref.~\cite{Silvestre-Brac:1996myf}}
\begin{tabular}{ c  c  c  c c c c c c c}
\hline\hline
$~~\kappa~~$& $~~\lambda~~$ & $~~\Lambda~~$& $~~\kappa^\prime~~$ &$~~A~~$&$~~B~~$& $~~m_n~~$& $~~m_s~~$&$~~m_c~~$&$~~m_b~~$ 
\\& $~~~~(\rm{GeV}^{5/3})~~~~$&$~~~~(\rm{GeV})~~~~$&& $~~~~(\rm{GeV}^{B-1})~~~~$&&$~~~~(\rm{GeV})~~~~$&$~~~~(\rm{GeV})~~~~$&$~~~~(\rm{GeV})~~~~$&$~~~~(\rm{GeV})~~~~$
\\ \hline
$0.4242$&$0.3898$&$1.1313$&$1.8025$& $1.5296$&$0.3263$&$0.277$&$0.553$&$1.819$&$5.206$
\\ \hline\hline
\end{tabular}
\label{tab:parameters}
\end{center}
\end{table*}

\begin{table}
\begin{center}
\renewcommand{\arraystretch}{1.5}
\caption{The masses and rms radii of heavy quarkonia and singly heavy mesons in the AP1 quark model. The experimental data are extracted from Ref.~\cite{ParticleDataGroup:2024cfk}. The “$\dagger$” represent possible experimental candidates discussed in Ref~\cite{Chen:2022asf}. }
\begin{tabular*}{\linewidth}{ c  c  c  c| c c c c}
\hline\hline
Meson& $m_{\rm{Exp}}$ & $m_{\rm{Theo}}$& $r_{\rm{Theo}}^{\rm{rms}}$ &Meson& $m_{\rm{Exp}}$ & $m_{\rm{Theo}}$& $r_{\rm{Theo}}^{\rm{rms}}$
\\ \hline
$\eta_c$&$2984$&$2983$&$0.35$& $\eta_b$&$9399$&$9403$&$0.20$
\\
$\eta_c(2S)$&$3638$&$3605$&$0.78$&$\eta_b(2S)$&$9999$&$10001$&$0.48$
\\
$J/\psi$&$3097$&$3103$&$0.40$&$\Upsilon$&$9460$&$9463$&$0.21$
\\
$\psi(2S)$&$3686$&$3646$&$0.81$&$\Upsilon(2S)$&$10023$&$10015$&$0.49$
\\
$D$&$1867$&$1882$&$0.65$&$\bar B$&$5279$&$5311$&$0.66$
\\ 
$D(2S)$&$2549^\dagger$&$2630$&$1.36$&$\bar B(2S)$&$5863^\dagger$&$6003$&$1.34$
\\
$D^*$&$2009$&$2033$&$0.75$&$\bar B^*$&$5325$&$5367$&$0.70$
\\
$D^*(2S)$&$2627^\dagger$&$2695$&$1.42$&$\bar B^*(2S)$&$5971^\dagger$&$6028$&$1.36$
\\ 
$D_s$&$1968$&$1957$&$0.50$&$\bar B_s$&$5367$&$5357$&$0.49$
\\
$D_s(2S)$&$2591^\dagger$&$2646$&$1.08$&$\bar B_s(2S)$&--&$5986$&$1.03$
\\
$D_s^*$&$2112$&$2108$&$0.58$&$\bar B_s^*$&$5415$&$5419$&$0.53$
\\
$D_s^*(2S)$&$2714^\dagger$&$2704$&$1.13$&$\bar B_s^*(2S)$&--&$6010$&$1.05$
\\
$B_c$&$6274$&$6269$&$0.30$&$\bar B_c$&$6274$&$6269$&$0.30$
\\ 
$B_c(2S)$&$6871$&$6854$&$0.66$&$\bar B_c(2S)$&$6871$&$6854$&$0.66$
\\
$B_c^*$&$6329$&$6338$&$0.32$&$\bar B_c^*$&$6329$&$6338$&$0.32$
\\
$B_c^*(2S)$&--&$6875$&$0.68$&$\bar B_c^*(2S)$&--&$6875$&$0.68$
\\ \hline\hline
\end{tabular*}
\label{tab:mass}
\end{center}
\end{table}

\subsection{Wave function}

The wave function $\Psi$ of the $S$-wave triply heavy tetraquark states can be expanded by the direct product of the spin wave function $\chi_s$, the color wave function $\chi_c$, and the spatial wave function $\phi$

\begin{eqnarray}
\Psi=\mathcal{A}(\chi_s\otimes \chi_c\otimes \phi)\, ,
\end{eqnarray}
where $\mathcal{A}$ is the antisymmetrization operator of the two identical quarks. When $Q_1=Q_2$, $\mathcal{A}=1-P_{12}$, otherwise $\mathcal{A}=1$, where $P_{ij}$ represents permutation operation on the $i$-th and $j$-th quarks.
 
\begin{figure}[H]
\begin{center}
\subfigure[]{
\scalebox{0.5}{\includegraphics{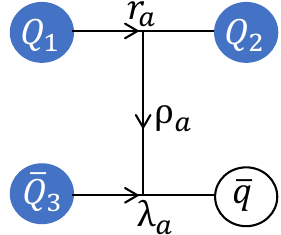}}}~~~~
\subfigure[]{
\scalebox{0.5}{\includegraphics{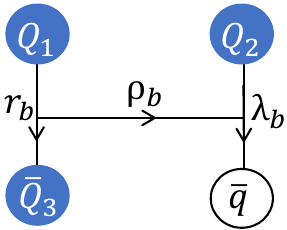}}}~~~~
\subfigure[]{
\scalebox{0.5}{\includegraphics{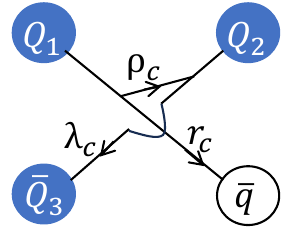}}}
\caption{The Jacobi coordinates for three types of spatial configurations: (a) for the diquark-antidiquark configuration, and (b), (c) for the dimeson configuration.}
\label{fig:Jacobi}
\end{center}
\end{figure}

We construct three sets of spatial configurations, including diquark-antidiquark cluster and dimeson cluster, which can be depicted by different Jacobi coordinates $r_\alpha$, $\lambda_\alpha$, and $\rho_\alpha$ as shown in Fig.~\ref{fig:Jacobi}. Especially, when triply heavy tetraquark system contains a pair of identical heavy quarks, the configuration in panel (b) is transformed into that in panel (c) due to the exchange of identical particles. Extra K-type coordinates are considered in Refs.~\cite{Meng:2020knc,Hiyama:2003cu}. 
% These configurations are not included in this work because quarks (antiquarks) carry color degrees of freedom and interact strongly with the remaining three-quark (antiquark) system at short range. K-type configurations are adequate and sometimes even necessary for four-body systems in QED or nuclear physics, where electromagnetic interactions or residual strong interactions occur between one particle and the remaining three at long range. For example, there exists scattering states composed of one free nucleon and the bound or resonant three-nucleon cluster. In contrast, there does not exist the free quark because of color confinement. 
These configurations are not included in this work due to color confinement. K-type configurations are adequate and sometimes even necessary for four-body systems in QED or nuclear physics, where electromagnetic interactions or residual strong interactions occur between one particle and the remaining three at long range. For example, in four-nucleon systems, there exist scattering states composed of one free nucleon and the three-nucleon cluster, which can be best described in K-type configurations. However, in tetraquark systems, such configurations do not exist because quarks (antiquarks) carry color degree of freedom, and color confinement prohibits the existence of free quark state.
Therefore, the K-type configurations can be omitted safely. We have verified that the current approach without K-type configurations already produces very precise results, as demonstrated in previous studies~\cite{Meng:2023jqk,Chen:2023syh,Ma:2024vsi,Wu:2024euj,Wu:2024hrv,Wu:2024zbx,Ma:2025rvj}. In this work, the spatial wave function $\phi_{nml}(r)$ adopts the Gaussian basis as

\begin{eqnarray}
\label{eq:gem}
\phi_{nml}&=&\sqrt{2^{l+5/2}\over \Gamma(l+3/2)r_n^3}({r\over r_n})^l e^{-{r^2\over r_n^2}}Y_{lm}(\hat{r})\, ,
\\ \nonumber r_n&=& r_0 a^{n-1}\, .
\end{eqnarray} 

The basis functions are not orthogonal but approximately complete if a large range of $r_n$ is considered. In principle, different choices of the sets of Jacobi coordinates should yield identical results if the basis functions are complete. To ensure completeness, one can select a set of Jacobi coordinates and construct the basis functions with total angular momentum $J$ by combining the spatial angular momenta of the three coordinates with the spin wave functions. However, managing angular momentum in GEM is a complex task, despite the existence of strategies for doing so~\cite{Hiyama:2003cu}. In this work, we only restrict the spatial wave functions to $l=0$ in Eq.~\ref{eq:gem}, while incorporating different Jacobi coordinates, which means that orbital excitation occurs when one set of Jacobi coordinates transforms to others. As shown in Fig.~\ref{fig:Jacobi}, we consider three different sets of Jacobi coordinates and use $n_{max}=14$ for each set, with $r_n$ following a geometric progression.

For color-spin wave functions with total angular momentum $J$, we choose a complete set of basis,

\begin{eqnarray}
\chi_1^{J=0}&=&[(Q_1 Q_2)_{\bar{3}_c}^{s_1=0}(\bar Q_3 \bar q/\bar s)_{3_c}^{s_2=0}]_{1_c}^{J=0}\, ,
\\  
\chi_2^{J=0}&=&[(Q_1 Q_2)_{\bar{3}_c}^{s_1=1}(\bar Q_3\bar q/\bar s)_{3_c}^{s_2=1}]_{1_c}^{J=0}\, ,
\\ 
\chi_3^{J=0}&=&[(Q_1 Q_2)_{6_c}^{s_1=0}(\bar Q_3 \bar q/\bar s)_{\bar{6}_c}^{s_2=0}]_{1_c}^{J=0}\, ,
\\ 
\chi_4^{J=0}&=&[(Q_1 Q_2)_{6_c}^{s_1=1}(\bar Q_3 \bar q/\bar s)_{\bar{6}_c}^{s_2=1}]_{1_c}^{J=0}\, ,
\\
\chi_5^{J=1}&=&[(Q_1 Q_2)_{\bar{3}_c}^{s_1=0}(\bar Q_3 \bar q/\bar s)_{3_c}^{s_2=1}]_{1_c}^{J=1}\, , 
\\  
\chi_6^{J=1}&=&[(Q_1 Q_2)_{\bar{3}_c}^{s_1=1}(\bar Q_3 \bar q/\bar s)_{3_c}^{s_2=0}]_{1_c}^{J=1}\, ,
\\ 
\chi_7^{J=1}&=&[(Q_1 Q_2)_{\bar{3}_c}^{s_1=1}(\bar Q_3 \bar q/\bar s)_{3_c}^{s_2=1}]_{1_c}^{J=1}\, ,
\\ 
\chi_8^{J=1}&=&[(Q_1 Q_2)_{6_c}^{s_1=0}(\bar Q_3 \bar q/\bar s)_{\bar{6}_c}^{s_2=1}]_{1_c}^{J=1}\, ,
\\ 
\chi_9^{J=1}&=&[(Q_1 Q_2)_{6_c}^{s_1=1}(\bar Q_3 \bar q/\bar s)_{\bar{6}_c}^{s_2=0}]_{1_c}^{J=1}\, ,
\\
\chi_{10}^{J=1}&=&[(Q_1 Q_2)_{6_c}^{s_1=1}(\bar Q_3 \bar q/\bar s)_{\bar{6}_c}^{s_2=1}]_{1_c}^{J=1}\, ,
\\
\chi_{11}^{J=2}&=&[(Q_1 Q_2)_{\bar{3}_c}^{s_1=1}(\bar Q_3\bar q/\bar s)_{3_c}^{s_2=1}]_{1_c}^{J=2}\, ,
\\ 
\chi_{12}^{J=2}&=&[(Q_1 Q_2)_{6_c}^{s_1=1}(\bar Q_3 \bar q/\bar s)_{\bar{6}_c}^{s_2=1}]_{1_c}^{J=2}\, .
\end{eqnarray}

We note that there is no definite $C$-parity in triply heavy tetraquark systems, so the overall wave  function has a simple form and only needs to satisfy the exchange antisymmetry when the system contains identical quarks. In Refs.~\cite{Ma:2024vsi,Wu:2024euj,Wu:2024hrv,Wu:2024ocq}, the more complicated overall wave functions of tetraquark states carry definite $C$-parity, which must be exactly positive or negative under the $C$-transformation.

\subsection{Complex scaling method}

The complex scaling method proposed by Aguilar, Balslev and Combes has been widely used to directly obtain the resonance energy and the decay width of resonances in atomic and molecular physics~\cite{Aguilar:1971ve,Balslev:1971vb,Aoyama:2006hrz}. In Refs.~\cite{Wang:2022yes,Chen:2023syh,Ma:2024vsi,Wu:2024euj,Wu:2024hrv,Wu:2024zbx}, the CSM was applied to obtain possible bound and resonant states in singly heavy, doubly heavy, and fully heavy tetraquark systems. We follow these works and conduct systematic calculations of $S$-wave triply heavy tetraquark states ($QQ^{(\prime)}\bar Q^{(\prime)}\bar q/\bar s$) with quantum number $J^P=0^+,1^+,2^+$. The CSM introduces the transformation $U(\theta)$ for the radial coordinate $r$ and its conjugate momentum $k$

\begin{eqnarray}
\nonumber U(\theta)r=r e^{i\theta}\, ,~~~~U(\theta)k=k e^{-i\theta}
\end{eqnarray}

The Hamiltonian is transformed as
\begin{eqnarray}
\nonumber H(\theta)&=& \sum_{i=1}^4\Big(m_i+{p_i^2e^{-2i\theta}\over 2m_i}\Big)+\sum_{i\textless j}V_{ij}(r_{ij}e^{i\theta})
\end{eqnarray}

Then we solve the complex-scaled four-body Schr\"{o}dinger equation. The wave function of $S$-wave tetraquark states with total angular momentum $J$ are expressed as
\begin{equation}
\Psi^J(\theta)=\mathcal{A}\sum_{\alpha,n_i,\beta}C_{\alpha,n_i,\beta}(\theta)\chi_\alpha^J\phi_{n_1}(r_\beta)\phi_{n_2}(\lambda_\beta)\phi_{n_3}(\rho_\beta)
\end{equation}
where $C_{\alpha,n_i,\beta}(\theta)$ is the complex expansion coefficient of basis function, which can be determined by solving the energy eigenvalue equation. The spatial wave function $\phi_{n_i}(r_\beta)$ is a real-range Gaussian basis function and does not undergo the operator $U(\theta)$ transformation, while the scaling angle $\theta$ is absorbed into the eigenvalue $E(\theta)$, eigenvector element $C_{\alpha,n_i,\beta}(\theta)$ and Hamiltonian $H(\theta)$. We write the complex-scaled Schr\"{o}dinger equation as
\begin{eqnarray}
H(\theta)\Psi^J(\theta)&=& E(\theta)\Psi^J(\theta)
\end{eqnarray}

According to the ABC theorem~\cite{Aguilar:1971ve,Balslev:1971vb}, the eigenenergies of scattering states, bound states, and resonant states can be determined by solving the complex-scaled Schr\"{o}dinger equations. The continuum states are arranged along rays originating from the threshold energies, with $\operatorname{Arg}(E)=2\theta$. The bound states are located on the negative real axis in the energy plane. Resonant states, characterized by mass $M_R$ and width $\Gamma_R$, are identified at $E_R=M_R-i\Gamma_R/2$, provided the complex scaling angle satisfies $2\theta\geq|\operatorname{Arg}(E_R)|$. Both bound and resonant states remain stable as the scaling angle $\theta$ is varied.

\subsection{Spatial structure}

\begin{figure}[H]
\begin{center}
\scalebox{0.8}{\includegraphics{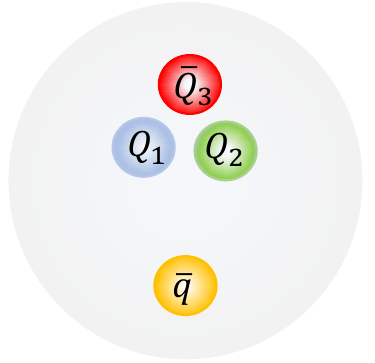}}
\end{center}
\caption{A typical spatial arrangement in triply heavy tetraquarks.}
\label{fig:triply}
\end{figure}

In addition to fundamental physical properties such as mass and width, the complex spatial structure constitutes an essential feature for characterizing the enigmatic nature of tetraquark states. To systematically analyze the spatial configurations of triply heavy tetraquark systems, we employ methodologies from Refs.~\cite{Ma:2024vsi,Wu:2024euj,Wu:2024hrv,Wu:2024zbx} to investigate their spatial distributions. Tetraquark states are broadly classified into two categories: meson molecules and compact tetraquarks. In the molecular configuration, the relative separation between two quark-antiquark clusters exceeds the characteristic scale of color confinement ($\Lambda_{QCD}^{-1}\sim 1$ fm). In contrast, compact tetraquarks involve confined (anti)quarks within a spatial region on the order of $\Lambda_{QCD}^{-1}$. A typical compact spatial arrangement in triply heavy tetraquarks is the triquark-centered configuration, where three heavy quarks cluster densely while a light quark orbits the triquark core, as illustrated in Fig.~\ref{fig:triply}. This structure exhibits an analogy to the QED-based tritium atom model. For brevity, we denote compact tetraquarks and molecular configurations as C. and M., respectively.

The root-mean-square (rms) radius is a direct measure to distinguish between different spatial configurations of tetraquark states. In the Ref.~\cite{Chen:2023syh}, Chen and Wu et al. proposed a new approach to calculate the rms radii in tetraquark systems. Briefly, we adopt the conventional definition of the rms radius when there are no identical particles in triply heavy tetraquark system (e.g., $c b\bar b \bar q $, $b c\bar c\bar q $, $c b\bar b \bar s $, $b c\bar c\bar s $). The conventional definition of the rms radius is

\begin{eqnarray}
r_{ij}^{rms,C}&=& \operatorname{Re}\Big[\sqrt{{\langle \Psi(\theta)|r^2_{ij}e^{2i\theta}|\Psi(\theta)\rangle\over \langle\Psi(\theta)|\Psi(\theta)\rangle}}\Big]\, .
\end{eqnarray}

However, when identical particles are present in the tetraquark state, the conventional rms radius is not suitable for describing the molecular configuration due to the antisymmetrization requirement for identical particles. Therefore, we use a new approach to calculate the rms radius. For systems containing one pair of identical particles (e.g., $cc\bar c\bar q $, $cc\bar b\bar q $, $bb\bar b\bar q $, $bb\bar c\bar q $, $cc\bar c\bar s $, $cc\bar b\bar s $, $bb\bar b\bar s $, $bb\bar c\bar s $), we decompose the color part of complete antisymmetric wave function as 
\begin{eqnarray}
\Psi(\theta)&=&\sum_{s_1,s_2}\Big[[(Q_1\bar Q_3)_{1_c}^{s_1}(Q_2\bar q)_{1_c}^{s_2}]_{1_c}^J\otimes \psi(r_1,r_2,r_3,r_4;\theta)\nonumber
\\ \nonumber &&-[(Q_2\bar Q_3)_{1_c}^{s_1}(Q_1\bar q)_{1_c}^{s_2}]_{1_c}^J\otimes \psi(r_2,r_1,r_3,r_4;\theta)\Big]
\\ \nonumber &=&\mathcal{A}\left[\sum_{s_1,s_2}[(Q_1\bar Q_3)_{1_c}^{s_1}(Q_2\bar q)_{1_c}^{s_2}]_{1_c}^J\otimes \psi(r_1,r_2,r_3,r_4;\theta)\right]
\\ &\equiv&\mathcal{A}\Psi_{nA}(\theta)\, .
\end{eqnarray}

Then we use the decomposed non-antisymmetric wave function $\Psi_{nA}(\theta)$ to calculate the rms radius
\begin{eqnarray}
r_{ij}^{rms,M}&=& \operatorname{Re}\Big[\sqrt{{\langle \Psi_{nA}(\theta)|r^2_{ij}e^{2i\theta}|\Psi_{nA}(\theta)\rangle\over \langle\Psi_{nA}(\theta)|\Psi_{nA}(\theta)\rangle}}\Big]\, .
\end{eqnarray}

The inner products in the CSM are defined by the c-product as
\begin{eqnarray}
\langle \phi_n|\phi_m\rangle=\int \phi_n(r) \phi_m (r)d^3 r
\end{eqnarray}

\section{Results and discussion}
\label{sec:results}

\subsection{$QQ\bar Q\bar q$}

\begin{figure*}[hbtp]
\begin{center}
\scalebox{0.6}{\includegraphics{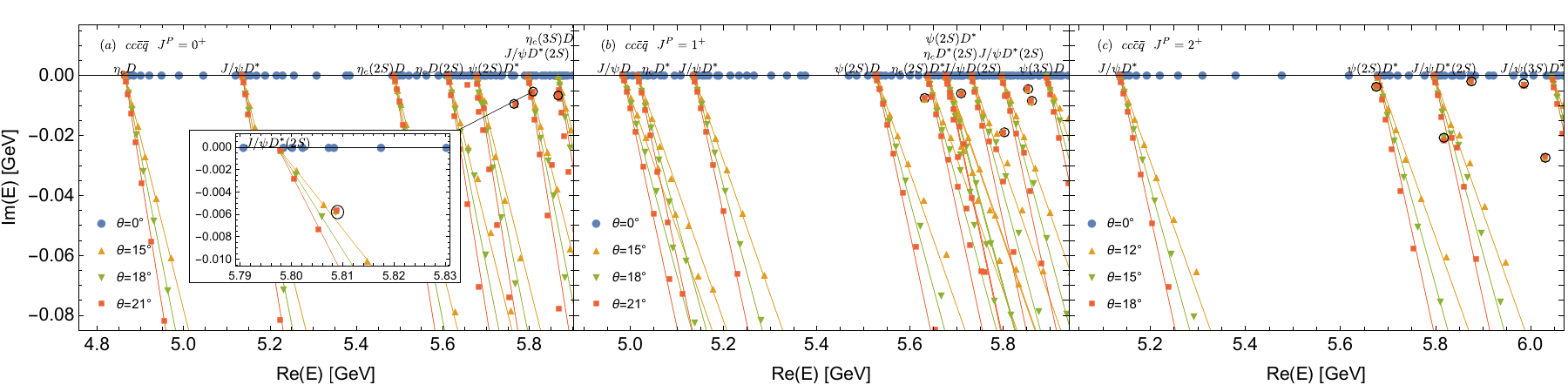}}
\end{center}
\caption{The complex eigenenergies of the $cc\bar c\bar q$ tetraquark states in the AP1 quark potential model with varying $\theta$ in the CSM. The solid lines represent the continuum lines rotating along $Arg(E)=-2\theta$. The resonant states remain stable as the scaling angle $\theta$ changes and marked out by the black circles. }
\label{fig:cccn}
\end{figure*}

\begin{figure*}[hbtp]
\begin{center}
\scalebox{0.6}{\includegraphics{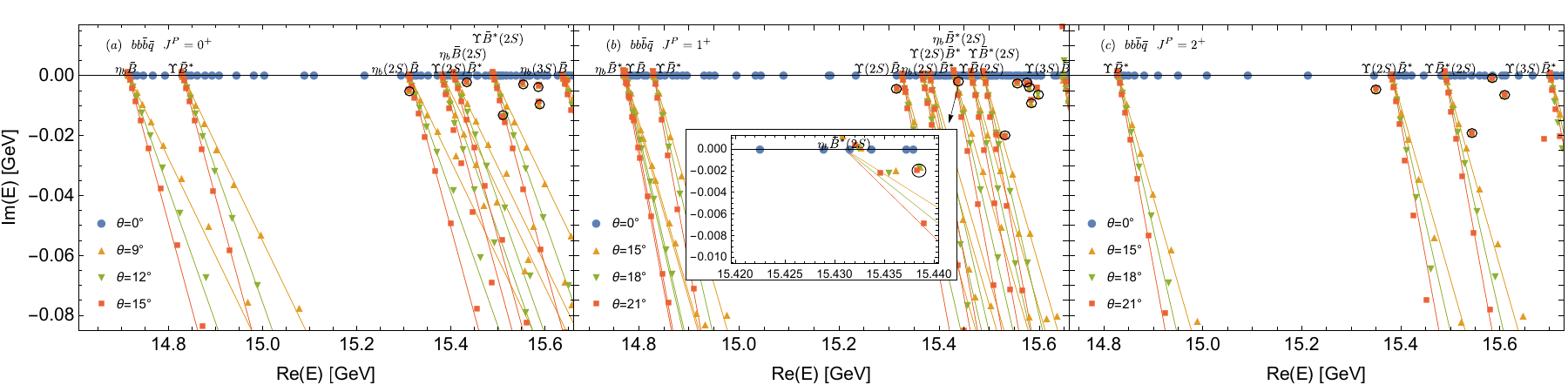}}
\end{center}
\caption{The complex eigenenergies of the $bb\bar b\bar q$ tetraquark states in the AP1 quark potential model with varying $\theta$ in the CSM.}
\label{fig:bbbn}
\end{figure*}

We employ the CSM to calculate the complex eigenenergies of the $S$-wave $QQ\bar Q \bar q$ systems, including states with quantum numbers $J^{P}=0^+$, $1^+$ and $2^+$. The distribution of complex eigenenergies for $cc\bar c\bar q$ and $bb\bar b\bar q$ systems is shown in  Figs.~\ref{fig:cccn}-\ref{fig:bbbn}, respectively. These figures exhibit several consistent features. For example, the continuum states are located along the continuum lines, rotated clockwise by $2\theta$ from the positive real axis. The origin of the continuum lines on the real axis corresponds to the thresholds of the di-meson states. Additionally, there are states situated between the positive real axis and the continuum lines, corresponding to poles on the second Riemann sheet of the respective thresholds. These states do not shift with $\theta$ and are identified as resonance states. No signal for bound states is found in the triply heavy tetraquark systems, consistent with the findings in Refs.~\cite{Lu:2021kut,Yang:2024nyc}. In the following discussion, we denote these resonances as $T_{3Q,J}(M)$, where $J$ represents the total angular momentum and $M$ is the mass of the state. The complex eigenenergies, proportions of color configurations and rms radii of the resonant states are listed in Table~\ref{tab:QQQq}.

\begin{table*}[hbtp]
\begin{center}
\renewcommand{\arraystretch}{1.5}
\caption{The complex eigenenergies $E=M-i{\Gamma/2}$, proportions of two color components, and rms radii between different quarks of the $S$-wave $QQ\bar Q\bar q$ tetraquark states. The last column indicates the spatial configuration of each state, where compact tetraquarks and molecular configurations are denoted as C. and M., respectively. The “?” signifies that the rms radii between quarks are numerically unstable as the complex scaling angle $\theta$ changes.}
\begin{tabular}{c c  c  c  c  c c  c  c c c c}
\hline\hline
system &$I(J^P)$&~~~$M-i\Gamma/2$ ~~~&~~~$\chi_{\bar{3}_c\otimes 3_c}$~~~&~~~$\chi_{6_c\otimes \bar{6}_c}$~~~&~~~$r_{Q_1\bar Q_3}^{\rm{rms}}$~~~&~~~$r_{Q_2\bar q}^{\rm{rms}}$~~~&~~~$r_{Q_1 Q_2}^{\rm{rms}}$~~~&~~~$r_{\bar Q_3\bar q}^{\rm{rms}}$~~~&~~~$r_{Q_1 \bar q}^{\rm{rms}}$~~~&~~~$r_{Q_2\bar Q_3}^{\rm{rms}}$~~~&Configuration
\\  \hline
$cc\bar c\bar q$&$1/2(0^+)$& $5768-10i$&$26\%$&$74\%$&$0.63$&$1.08$&$0.75$&$1.17$&$1.22$&$0.79$& C.
\\
&&$5809-6i$&$77\%$&$23\%$&$0.66$&$1.03$&?&?&?&?&?
\\
&&$5867-7i$&$54\%$&$46\%$&$0.58$&$1.15$&$0.80$&$1.14$&$1.15$&$0.72$&C.
\\
&$1/2(1^+)$&$5634-8i$&$81\%$&$19\%$&$0.82$&$0.91$&$0.73$&$0.98$&$1.04$&$0.95$&C.
\\
&&$5711-6i$&$39\%$&$61\%$&$0.68$&$0.91$&$0.81$&$0.98$&$0.99$&$0.88$&C.
\\
&&$5801-19i$&$23\%$&$77\%$&$0.73$&$0.88$&$1.03$&$1.14$&$1.29$&$0.81$&C.
\\
&&$5853-5i$&$63\%$&$37\%$&$0.59$&$1.11$&$0.79$&$1.10$&$1.10$&$0.75$&C.
\\
&&$5861-9i$&$52\%$&$48\%$&$0.59$&$1.13$&$0.76$&$1.22$&$1.22$&$0.72$&C.
\\
&$1/2(2^+)$&$5677-4i$&$73\%$&$27\%$&$0.84$&$0.89$&$0.97$&$1.17$&$1.19$&$1.11$&C.
\\ 
&&$5819-20i$&$27\%$&$73\%$&$0.69$&$0.97$&$2.08$&$2.11$&$2.20$&$2.03$&M.
\\
&&$5876-2i$&$71\%$&$29\%$&$0.61$&$1.09$&$0.81$&$1.06$&$1.06$&$0.71$&C.
\\
&&$5987-3i$&$91\%$&$9\%$&$1.10$&$1.21$&$0.61$&$0.99$&$1.22$&$1.14$&C.
\\
&&$6031-28i$&$87\%$&$13\%$&$0.86$&$1.13$&$0.54$&$1.22$&$1.11$&$0.76$&C.
\\ \hline
$bb\bar b\bar q$&$1/2(0^+)$& $15312-5i$&$73\%$&$27\%$&$0.52$&$0.75$&$0.41$&$0.75$&$0.75$&$0.52$& C.
\\
&&$15435-2i$&$45\%$&$55\%$&$0.45$&$0.82$&$0.49$&$0.83$&$0.81$&$0.60$&C.
\\
&&$15511-14i$&$47\%$&$53\%$&?&?&?&?&?&?&?
\\
&&$15554-3i$&$60\%$&$40\%$&$0.39$&$0.99$&$0.57$&$0.98$&$0.99$&$0.55$&C.
\\
&&$15587-3i$&$77\%$&$23\%$&$0.65$&$0.93$&$0.46$&$0.89$&$0.95$&$0.67$&C.
\\
&&$15588-9i$&$58\%$&$42\%$&$0.57$&$1.03$&$0.41$&$0.96$&$1.01$&$0.60$&C.
\\
&$1/2(1^+)$& $15316-4i$&$80\%$&$20\%$&$0.53$&$0.77$&$0.49$&$0.84$&$0.86$&$0.65$& C.
\\ 
&&$15438-2i$&$38\%$&$62\%$&$0.44$&$0.83$&$0.50$&$0.84$&$0.82$&$0.60$&C.
\\
&&$15533-21i$&$46\%$&$54\%$&$0.46$&$0.84$&$0.64$&$0.95$&$1.04$&$0.56$&C.
\\
&&$15558-2i$&$56\%$&$44\%$&$0.39$&$1.01$&$0.49$&$0.96$&$0.97$&$0.48$&C.
\\
&&$15576-3i$&$48\%$&$52\%$&$0.33$&$1.03$&$0.50$&$1.03$&$1.02$&$0.45$&C.
\\
&&$15582-4i$&$72\%$&$28\%$&$0.59$&$0.97$&$0.46$&$0.91$&$0.97$&$0.63$&C.
\\
&&$15586-9i$&$67\%$&$33\%$&$0.64$&$1.00$&$0.42$&$0.91$&$1.01$&$0.68$&C.
\\
&&$15599-6i$&$92\%$&$8\%$&$0.74$&$0.92$&$0.40$&$0.81$&$0.95$&$0.75$&C.
\\
&$1/2(2^+)$& $15351-5i$&$88\%$&$12\%$&$0.54$&$0.80$&$0.39$&$0.81$&$0.84$&$0.61$& C.
\\ 
&&$15544-19i$&$49\%$&$51\%$&$0.48$&$0.81$&$0.57$&$0.89$&$0.98$&$0.53$&C.
\\
&&$15585-1i$&$52\%$&$48\%$&$0.37$&$1.03$&$0.50$&$1.01$&$1.01$&$0.47$&C.
\\
&&$15611-7i$&$91\%$&$9\%$&$0.75$&$0.94$&$0.38$&$0.82$&$0.97$&$0.75$&C.
\\ \hline\hline
\end{tabular}
\label{tab:QQQq}
\end{center}
\end{table*}

\subsubsection{$cc\bar c \bar q$}

For the $S$-wave $cc\bar c\bar q$ system, we identify thirteen resonant states with quantum numbers $J^P=0^+$, $1^+$ and $2^+$. We focus on the states that lie below the $M_1(3S)M_2(1S)$ threshold, with masses ranging from $5.7$~GeV to $6.1$~GeV. Three resonant states with $J^P=0^+$ are identified: $T_{3c,0}(5768)$, $T_{3c,0}(5809)$, and $T_{3c,0}(5867)$. They have same decay channels: $\eta_c D$, $J/\psi D^*$, $\eta_c(2S)D$, $\eta_c D(2S)$ and $\psi(2S) D^*$. The $T_{3c,0}(5768)$, with a width of $\Gamma=20$~ MeV, has a compact configuration, where the charm quarks are close to each others while the light quark is slightly further away, resembling a tritium-like structure. The $\chi_{6_c\otimes \bar 6_c}$ color component of this state is up to $74\%$, indicating a strong attractive interaction between diquark and antidiquark clusters. The rms radii of $T_{3c,0}(5809)$ exhibit numerical instability and significant variations with changes in $\theta$. This instability arises because the $T_{3c,0}(5809)$ can strongly couple to the $J/\psi D^*(2S)$ continuum states located on the $2\theta$-line. Consequently, more precise calculations are needed to obtain an accurate spatial distribution for this state. The $T_{3c,0}(5867)$ is a compact resonant state with a width of $\Gamma=14$~MeV. It has approximately the same color composition of $\chi_{\bar 3_c\otimes 3_c}$ and $\chi_{6_c\otimes \bar 6_c}$, and can also decay into $J/\psi D^*(2S)$ channel.

We also obtain five resonant states with spin-parity $J^P=1^+$: $T_{3c,1}(5634)$, $T_{3c,1}(5711)$, $T_{3c,1}(5801)$, $T_{3c,1}(5853)$, and $T_{3c,1}(5861)$, all of which have a compact tetraquark configuration. The $T_{3c,1}(5634)$, with a width of $\Gamma=16$~MeV, is the lowest triply heavy tetraquark resonant sate. It contains an $81\%$ $\chi_{\bar 3_c\otimes 3_c}$ component and can be observed in experiments through the decay channels $J/\psi$, $\eta_c D^*$, $J/\psi D^*$, and $\psi(2S) D$. The spatial distribution of this state is nearly uniform, where the rms radii are approximately the same, with the two identical charm quarks positioned closest to each other. The higher state $T_{3c,1}(5711)$ is located above the $\psi(2S)D^*$ threshold, with a width of $\Gamma=12$~MeV. It has three additional decay channels: $\eta_c(2S)D^*$, $\eta_cD^*(2S)$, and $\psi(2S)D^*$. The states $T_{3c,1}(5853)$ and $T_{3c,1 }(5961)$ have a mass splitting of only $8$~MeV and possess similar color components, rms radii, and decay widths, which makes it difficult to distinguish between them.

Additionally, we investigate the mass spectra, decay channels, color configurations and rms radii of five resonant states with quantum number $J^P=2^+$: $T_{3c,2}(5677)$, $T_{3c,2}(5819)$, $T_{3c,2}(5876)$, $T_{3c,2}(5987)$, and $T_{3c,2}(6031)$. The compact tetraquark state $T_{3c,2}(5677)$ exhibits only one strong decay channel, $J/\psi D^*$, and could be experimentally probed via $J/\psi D\pi$ final states. The higher state $T_{3c,2}(5819)$ is a broader resonant state with a width of $\Gamma=40$~MeV. Notably, the rms radii $r_{c_1\bar c_3}^{\rm rms}$ and $r_{c_2\bar q}^{\rm rms}$ are significantly smaller than other radii, being slightly larger than those of $J/\psi $ and $D^*$ and smaller than those of  $\psi(2S)$ and $D^*(2S)$. This suggests that this state may have molecular components. We recommend experimental exploration of this state in $J/\psi D^*$ and $\psi(2S) D^*$ decay channels. Furthermore, the compact state $T_{3c,2}(5987)$ has a substantial $\chi_{\bar 3_c\otimes 3_c}$ color configuration, up to $91\%$, indicating that the spatial wave function of the identical particles is largely symmetric.

\subsubsection{$bb\bar b \bar q$}

For the $S$-wave $bb\bar b \bar q$ system, we identify eighteen resonant states with quantum numbers $J^P=0^+$, $1^+$, and $2^+$. We first focus on the tetraquarks with $J^P=0^+$ that lie below the $\eta_b(3S)\bar B$ threshold. These include six resonant states: $T_{3b,0}(15312)$, $T_{3b,0}(15435)$, $T_{3b,0}(15511)$, $T_{3b,0}(15554)$, $T_{3b,0}(15587)$, and $T_{3b,0}(15588)$. Except $T_{3b,0}(15511)$, all of these states share common decay channels, such as $\eta_b\bar B$ and $\Upsilon \bar B^*$, and exhibit a compact spatial configuration where the three bottom quarks are relatively close to one another, while the light quarks are more distant. The lowest state $T_{3b,0}(15312)$, with a width of $\Gamma=10$~MeV, is characterized by a dominant $\chi_{\bar 3_c\otimes 3_c}$ color configuration. It is more likely to be observed in $S$-wave $\eta_b\bar B$ and $P$-wave $\Upsilon \bar B \pi$ final states. The higher state $T_{3b,0}(15435)$ has additional decay channels, including $\eta_b(2S) \bar B$, $\Upsilon(2S) \bar B^*$, and $\eta_b \bar B(2S)$. The state $T_{3b,0}(15554)$ may correspond to the bottom partner state of $T_{3c,0}(5867)$. The mass splitting between these states is approximately $3(m_b-m_c)$. Furthermore, they exhibit similar color configurations and semblable spatial distribution. Specifically, the rms radius $r_{Q_1\bar Q_3}$ is the smallest among all the radii, while the $r_{Q_1 Q_2}$ and $r_{Q_2\bar Q_3}$ are comparable, and the remaining radii are roughly equal. Another intriguing structure, $T_{3b,0}(15587)$, has a dominant $\chi_{\bar 3_c\otimes 3_c}$ color component (up to $77\%$), which is similar to that of the lowest state, $T_{3b,0}(15312)$. The rms radii between quarks for this state are slightly larger than those of $T_{3b,0}(15312)$. This suggest that $T_{3b,0}(15587)$ could potentially serve as the radial excitation of $T_{3b,0}(15312)$. Additionally, $T_{3b,0}(15587)$ has extra decay channels, including $\eta_b(2S)\bar B$, $\Upsilon(2S)\bar B^*$, $\eta_b \bar B(2S)$ and $\Upsilon\bar B^*(2S)$.

In the sector with quantum number $J^P=1^+$, we obtain eight compact resonant states below the $\Upsilon(3S) \bar B$ threshold. Their masses range from $15.3$ to $15.6$~GeV, and all share the same decay channels: $\eta_b \bar B^*$, $\Upsilon \bar B$, and $\Upsilon \bar B^*$. The states $T_{3b,1}(15438)$ may correspond to the bottom partner state of $T_{3c,1}(5711)$, as they both lie below the corresponding threshold of heavy quarkonium and heavy meson. Their mass difference is also approximately $3(m_b-m_c)$. Moreover, the state $T_{3b,1}(15438)$ can decay into additional final states, including $\eta_b(2S)\bar B^*$, $\Upsilon(2S)\bar B^*$ and $\eta_b\bar B^*$. The states $T_{3b,1}(15558)$ and $T_{3b,1}(15576)$ have more decay channels than $T_{3b,1}(15438)$, including $\Upsilon \bar B(2S)$ and $\Upsilon \bar B^*(2S)$.

We also identify four compact resonant states in the $bb\bar b \bar q$ system with quantum number $J^P=2^+$. Their mass spectra range  from $15.3$ to $15.7$~GeV, below the threshold of $\Upsilon(3S) \bar B^*$. The color-spin configuration $[(bb)_{\bar 3_c}^1(\bar b \bar q)_{3_c}^1]_{1c}^2$ is predominant component in wave functions of $T_{3b,2}(15351)$ and $T_{3b,2}(15611)$, which are similar to the color-spin compositions of $T_{3c,2}(5677)$ and $T_{3c,2}(5987)$. We treat these states as bottom-charm partners, respectively.
%, with a mass difference of approximately $9.7$~GeV.
Notably, the state $T_{3b,2}(15351)$ may be more likely to be observed in $\Upsilon \bar B \pi$ final states in experiments. In contrast, the state $T_{3b,2}(15611)$ has additional decay channels, including $\Upsilon(2S) \bar B^*$ and $\Upsilon \bar B^*(2S)$.

\subsection{$QQ\bar Q^\prime \bar q$}

\begin{figure*}[hbtp]
\begin{center}
\scalebox{0.6}{\includegraphics{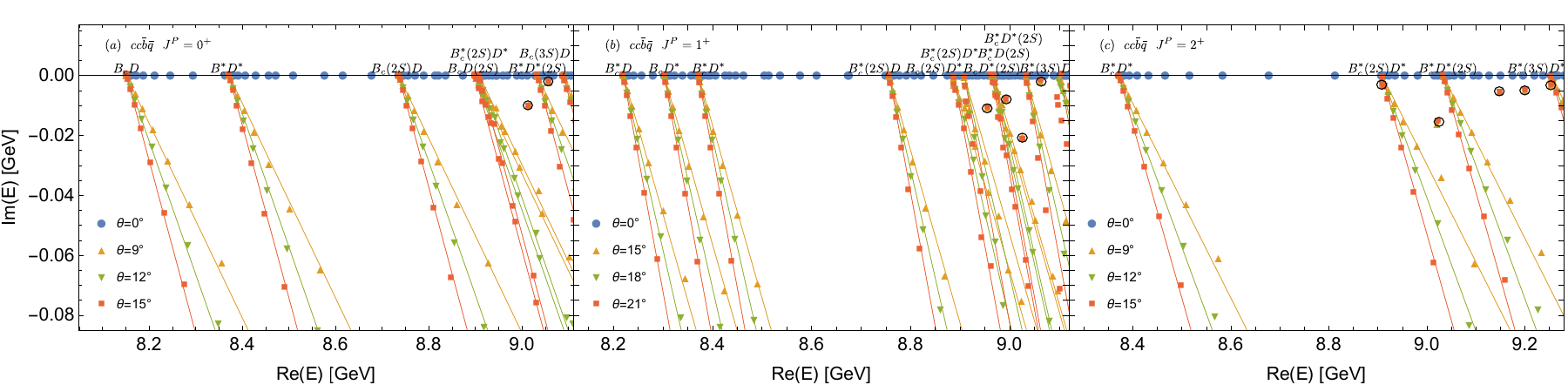}}
\end{center}
\caption{The complex eigenenergies of the $cc\bar b\bar q$ tetraquark states in the AP1 quark potential model with varying $\theta$ in the CSM.}
\label{fig:ccbn}
\end{figure*}

\begin{figure*}[hbtp]
\begin{center}
\scalebox{0.6}{\includegraphics{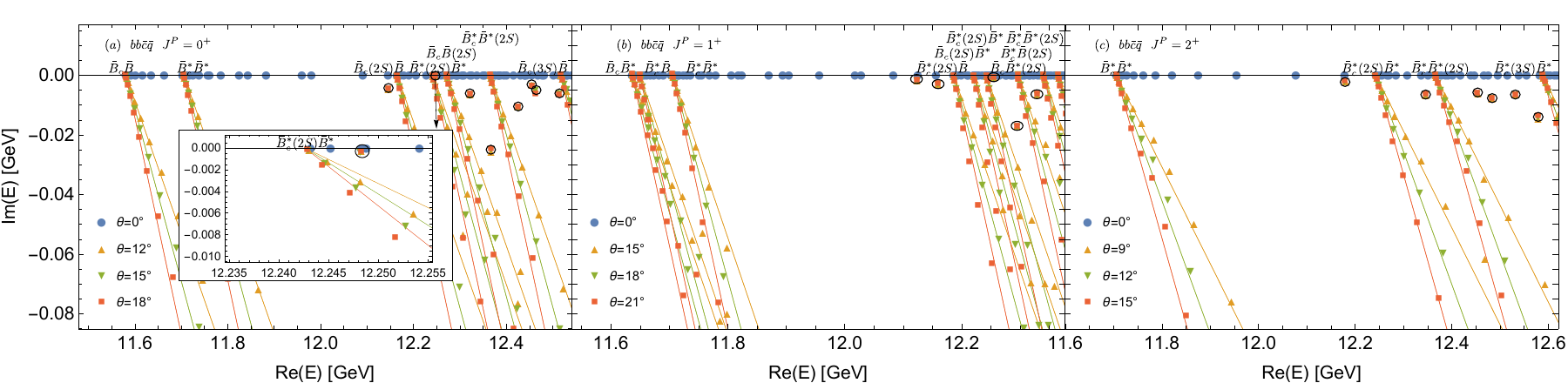}}
\end{center}
\caption{The complex eigenenergies of the $bb\bar c\bar q$ tetraquark states in the AP1 quark potential model with varying $\theta$ in the CSM.}
\label{fig:bbcn}
\end{figure*}

We investigate the complex eigenenergies of $S$-wave $QQ\bar Q^\prime\bar q$ systems within the CSM framework. For the $cc\bar b\bar q$ system, we obtain eleven resonant states with quantum numbers $J^P=0^+$, $1^+$, and $2^+$. While the number of resonant states in the $bb\bar c\bar q$ system is as high as nineteen. These resonant states are denoted as $T_{QQ Q^\prime,J}(M)$. The results for $cc\bar b\bar q$ and $bb\bar c\bar q$ systems are shown in Figs.~\ref{fig:ccbn}-\ref{fig:bbcn}. The complex eigenenergies, proportions of two color components and rms radii of different quarks are listed in Table~\ref{tab:QQQpq}.

\begin{table*}[hbtp]
\begin{center}
\renewcommand{\arraystretch}{1.5}
\caption{The complex eigenenergies, proportions of two color components, and rms radii between different quarks of the $S$-wave $QQ\bar Q^\prime \bar q$ tetraquark states. The last column indicates the spatial configuration of each state.}
\begin{tabular}{c c  c  c  c  c c  c  c c c c}
\hline\hline
system &$I(J^P)$&~~~$M-i\Gamma/2$ & $\chi_{\bar{3}_c\otimes 3_c}$~~~&~~~$\chi_{6_c\otimes \bar{6}_c}$~~~&~~~$r_{Q_1\bar Q_3}^{\rm{rms}}$~~~&~~~$r_{Q_2\bar q}^{\rm{rms}}$~~~&~~~$r_{Q_1 Q_2}^{\rm{rms}}$~~~&~~~ $r_{\bar Q_3\bar q}^{\rm{rms}}$~~~&~~~$r_{Q_1 \bar q}^{\rm{rms}}$~~~&~~~$r_{Q_2\bar Q_3}^{\rm{rms}}$~~~& Configuration
\\  \hline
$cc\bar b\bar q$&$1/2(0^+)$& $9015-10i$&$44\%$&$56\%$&$0.53$&$1.06$&$0.81$&$0.94$&$1.06$&$0.86$& C.
\\
&&$9057-2i$&$26\%$&$74\%$&$0.53$&$1.05$&$0.82$&$1.00$&$1.03$&$0.65$&C.
\\
&$1/2(1^+)$&$8956-11i$&$41\%$&$59\%$&$0.55$&$1.01$&$0.75$&$0.83$&$0.98$&$0.74$&C.
\\
&&$8993-8i$&$31\%$&$69\%$&$0.47$&$1.08$&$0.91$&$1.17$&$1.24$&$0.95$&C.
\\
&&$9026-21i$&$40\%$&$60\%$&$0.53$&$1.04$&$0.90$&$1.03$&$1.17$&$0.83$&C.
\\
&&$9064-2i$&$19\%$&$81\%$&$0.52$&$1.08$&$0.82$&$1.01$&$1.03$&$0.68$&C.
\\
&$1/2(2^+)$&$8909-3i$&$66\%$&$34\%$&$0.70$&$0.86$&?&?&?&?&?
\\ 
&&$9024-15i$&$25\%$&$75\%$&$0.51$&$1.06$&$0.80$&$0.87$&$1.06$&$0.67$&C.
\\
&&$9150-6i$&$91\%$&$9\%$&$0.63$&$1.02$&$0.84$&$0.84$&$1.01$&$0.65$&C.
\\
&&$9201-5i$&$89\%$&$11\%$&$0.89$&$1.13$&$0.70$&$0.89$&$1.14$&$0.98$&C.
\\
&&$9254-3i$&$43\%$&$57\%$&$0.68$&$1.18$&$0.86$&$1.03$&$1.14$&$0.97$&C.
\\ \hline
$bb\bar c\bar q$&$1/2(0^+)$& $12146-5i$&$85\%$&$15\%$&$0.72$&$0.81$&$0.45$&$0.92$&$0.88$&$0.79$& C.
\\
&&$12248-0.3i$&$61\%$&$39\%$&$0.60$&$0.83$&$0.58$&$0.97$&$0.90$&$0.74$&C.
\\
&&$12323-6i$&$74\%$&$26\%$&$0.58$&$0.85$&$0.55$&$0.81$&$0.83$&$0.54$&C.
\\
&&$12368-26i$&$57\%$&$43\%$&$0.49$&$1.08$&$0.45$&$0.68$&$0.73$&$0.44$&C.
\\
&&$12426-11i$&$71\%$&$29\%$&$0.84$&$1.06$&$0.40$&$1.12$&$1.11$&$0.88$&C.
\\
&&$12457-3i$&$52\%$&$48\%$&$0.82$&$1.06$&$0.48$&$1.06$&$1.10$&$0.85$&C.
\\
&&$12465-4i$&$45\%$&$55\%$&$0.60$&$1.06$&$0.65$&$1.14$&$1.06$&$0.70$&C.
\\
&&$12515-6i$&$74\%$&$26\%$&$0.69$&$1.02$&$0.56$&$1.13$&$1.04$&$0.69$&C.
\\
&$1/2(1^+)$& $12124-2i$&$94\%$&$6\%$&$0.73$&$0.84$&$0.35$&$0.83$&$0.86$&$0.76$& C.
\\ 
&&$12160-3i$&$91\%$&$9\%$&$0.73$&$0.83$&$0.39$&$0.90$&$0.87$&$0.78$&C.
\\
&&$12255-1i$&$47\%$&$53\%$&$0.58$&$0.85$&$0.55$&$0.98$&$0.88$&$0.71$&C.
\\
&&$12296-17i$&$69\%$&$31\%$&$0.57$&$0.84$&$0.62$&$0.79$&$0.86$&$0.48$&C.
\\
&&$12330-6i$&$72\%$&$28\%$&$0.57$&$0.84$&$0.55$&$0.82$&$0.83$&$0.53$&C.
\\
&$1/2(2^+)$& $12180-2i$&$93\%$&$7\%$&$0.74$&$0.85$&$0.36$&$0.89$&$0.88$&$0.78$& C.
\\ 
&&$12346-6i$&$70\%$&$30\%$&$0.57$&$0.88$&$0.58$&$0.87$&$0.87$&$0.56$&C.
\\
&&$12454-6i$&$88\%$&$12\%$&$0.92$&$1.07$&$0.36$&$1.06$&$1.08$&$0.96$&C.
\\
&&$12483-8i$&$45\%$&$54\%$&$0.71$&$1.13$&$0.38$&$1.16$&$1.17$&$0.74$&C.
\\
&&$12532-7i$&$69\%$&$31\%$&$0.66$&$1.04$&$0.54$&$1.14$&$1.05$&$0.66$&C.
\\
&&$12579-14i$&$65\%$&$35\%$&$0.69$&$1.01$&$0.76$&$1.02$&$0.96$&$0.70$&C.
\\ \hline\hline
\end{tabular}
\label{tab:QQQpq}
\end{center}
\end{table*}

Below the $B_c(3S)D$ threshold, there are two compact tetraquark states: $T_{ccb,0}(9015)$ and $T_{ccb,0}(9057)$. Both of them can decay into several channels, including $B_c D$, $B_c^* D^*$, $B_c(2S) D$, $B_c D(2S)$ and $B_c^*(2S)D^*$. The sates $T_{ccb,0}(9015)$ has a width of approximately $\Gamma=20$~MeV, while $T_{ccb,0}(9057)$ has a width of about $\Gamma=4$~MeV. In the $S$-wave $cc\bar b\bar q$ system with spin-parity $J^P=1^+$, four compact resonant states, $T_{ccb,1}(8956)$, $T_{ccb,1}(8993)$, $T_{ccb,1}(9026)$, and $T_{ccb,1}(9064)$, lie below the $B_c^*(3S)D$ threshold. Among these, The lowest state, $T_{ccb,1}(8956)$, has a relatively wide width of $\Gamma=22$~MeV and is more likely to be experimentally observed in decay channels $B_c^* D$, $B_cD^*$, $B_c^*D^*$, $B_c^*(2S)D$, $B_c(2S)D^*$ and $B_c^*(2S)D^*$. Additionally, the $J^P=2^+$ $cc\bar b\bar q$ system includes five resonant states below the $B_c^*(3S)D^*$ threshold. These states have masses ranging from $8.9$ to $9.3$~ GeV. The state $T_{ccb,2}(8909)$ is the lowest tetraquark in $cc\bar b\bar q$ system, with a width of $\Gamma=6$~MeV. The color-spin composition of this state, $[(cc)_{\bar 3_c}^1(\bar b\bar q)_{3_c}^1]_{1_c}^2$, is $66\%$, and the rms radii $r_{c_1\bar b}^{\rm rms}$ and $r_{c_2\bar q}^{\rm rms}$ are comparable to the radii of $B_c^*(2S)$ and $D^*$. However, due to the proximity of this state to the continuum states, other radii, such as $r_{c_1 c_2}^{\rm rms}$, $r_{\bar b\bar q}^{\rm rms}$, $r_{c_1\bar q}^{\rm rms}$ and $r_{c_2\bar b}^{\rm rms}$, are numerically less stable in CSM. Additionally, the higher resonant states $T_{ccb,2}(9024)$, $T_{ccb,2}(9150)$, $T_{ccb,2}(9201)$, and $T_{ccb,2}(9254)$ have compact configurations.

For the $S$-wave $bb\bar c \bar q$ system, we obtain eight compact resonant states with $J^P=0^+$. These states have mass spectra in the range of $12.1$-$12.6$~GeV and they share the same decay channels, including $\bar B_c\bar B$ and $\bar B_c^*\bar B^*$. Among them, the lowest state $T_{bbc,0}(12146)$ lies below $\bar B_c(2S)\bar B$ threshold and has a width of $\Gamma=10$ MeV. The primary color configuration for this state is $\chi_{\bar 3_c\otimes 3_c}$, and the rms radius between two bottom quarks is the smallest. The state $T_{bbc,0}(12248)$ has two additional decay channels: $\bar B_c(2S)\bar B$ and $\bar B_c^*(2S) \bar B^*$. However, the width of this state is nearly zero, suggesting that the decays are suppressed.  The suppression may arise from the following reasons: a) The overlap between initial and final spatial wave functions could be very small, leading to an extremely narrow width; b) The color mixing elements via color-electric terms between the initial and final states might be suppressed as discussed in Ref.~\cite{Wu:2024euj}, which vanish if the coefficients of color factor $\lambda_i\cdot \lambda_j$ are the the same; c) The almost-zero width of this state may result from its potential coupling to $P$-wave dimeson channels, which is not considered in our calculations. Since we do not calculate the partial width, the exact reasons need further analysis. In the following discussions, we also obtain several resonances with very narrow widths in other systems. 

Additionally, We identify five resonant states with $J^P=1^+$: $T_{bbc,1}(12124)$, $T_{bbc,1}(12160)$, $T_{bbc,1}(12255)$, $T_{bbc,1}(12296)$, and $T_{bbc,1}(12330)$. These states have compact tetraquark configurations and share the common decay channels $\bar B_c\bar B^*$, $\bar B_c^*\bar B$, and $\bar B_c^*\bar B^*$. The state $T_{bbc,1}(12124)$ is the lowest-lying resonant state in the $bb\bar c\bar q$ system, with a $94\%$ $\chi_{\bar 3_c\otimes 3_c}$ component. For the $bb\bar c\bar q$ with quantum number $J^P=2^+$, we acquire six compact resonant states: $T_{bbc,2}(12180)$, $T_{bbc,2}(12346)$, $T_{bbc,2}(12454)$, $T_{bbc,2}(12483)$, $T_{bbc,2}(12532)$, and $T_{bbc,2}(12579)$. The state $T_{bbc,2}(12180)$ has a dominant color configuration of $\chi_{\bar 3_c\otimes 3_c}$. The spatial wave function between two bottom quarks must be symmetric. It has a width of $\Gamma=4$ MeV and can be experimentally observed in the final states $\bar B_c^* (J/\psi l\nu_l)\bar B^*(\bar B\gamma)$.

\subsection{$QQ^\prime\bar Q^\prime\bar q$}

\begin{figure*}[hbtp]
\begin{center}
\scalebox{0.6}{\includegraphics{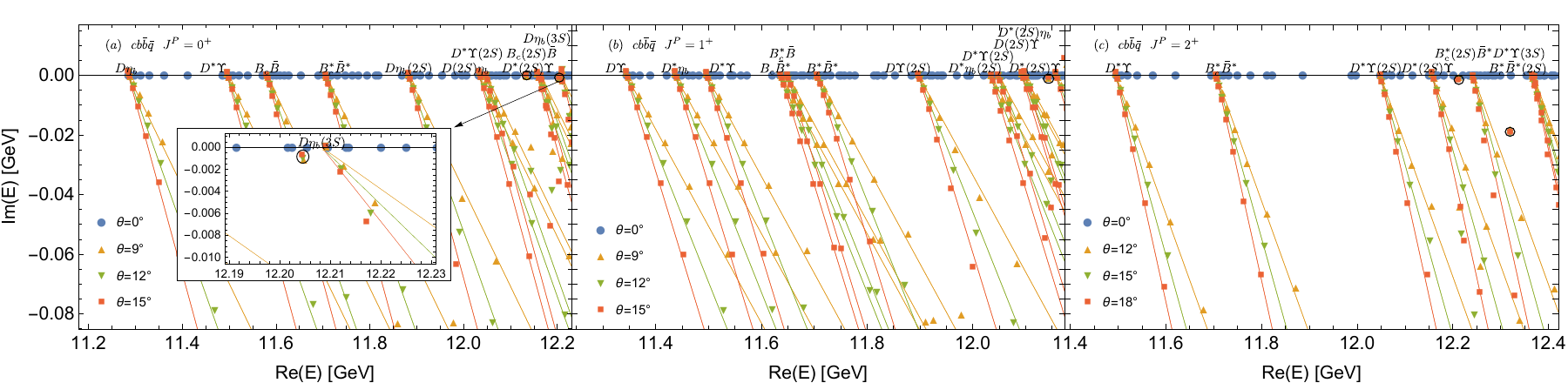}}
\end{center}
\caption{The complex eigenenergies of the $cb\bar b\bar q$ tetraquark states in the AP1 quark potential model with varying $\theta$ in the CSM.}
\label{fig:cbbn}
\end{figure*}

\begin{figure*}[hbtp]
\begin{center}
\scalebox{0.6}{\includegraphics{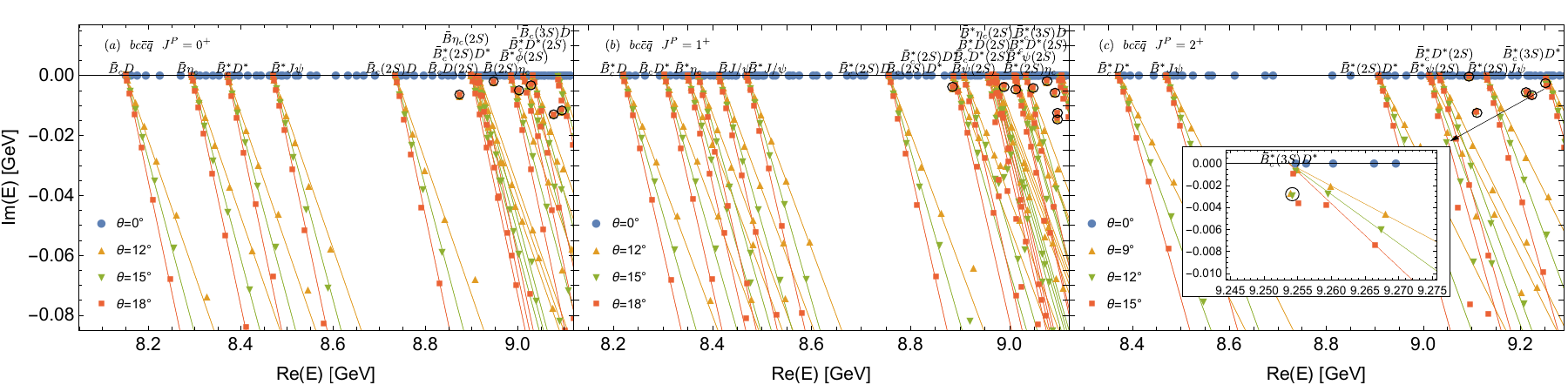}}
\end{center}
\caption{The complex eigenenergies of the $bc\bar c\bar q$ tetraquark states in the AP1 quark potential model with varying $\theta$ in the CSM.}
\label{fig:bccn}
\end{figure*}

We calculate the complex eigenenergies of the $S$-wave $QQ^\prime\bar Q^\prime\bar q$ system within the CSM framework and identify several resonant states, denoted as $T_{QQ^\prime Q^\prime,J}(M)$. The results for the $cb\bar b\bar q$ and $bc\bar c\bar q$ systems are shown in Figs.~\ref{fig:cbbn}-\ref{fig:bccn}. For these states, the complex eigenenergies, proportions of two color components and rms radii of different quarks are listed in Table~\ref{tab:QQpQpq}.

\begin{table*}[hbtp]
\begin{center}
\renewcommand{\arraystretch}{1.5}
\caption{The complex eigenenergies, proportions of two color components, and rms radii between different quarks of the $S$-wave $QQ^\prime\bar Q^\prime \bar q$ tetraquark states. The last column indicates the spatial configuration of each state.}
\begin{tabular}{c c  c  c  c  c c  c  c c c c}
\hline\hline
system &$I(J^P)$&~~~$M-i\Gamma/2$~~~&~~~$\chi_{\bar{3}_c\otimes 3_c}$~~~&~~~$\chi_{6_c\otimes \bar{6}_c}$~~~&~~~$r_{Q_1\bar Q_3}^{\rm{rms}}$~~~&~~~$r_{Q_2\bar q}^{\rm{rms}}$~~~&~~~$r_{Q_1 Q_2}^{\rm{rms}}$~~~&~~~ $r_{\bar Q_3\bar q}^{\rm{rms}}$~~~&~~~$r_{Q_1 \bar q}^{\rm{rms}}$~~~&~~~$r_{Q_2\bar Q_3}^{\rm{rms}}$~~~& Configuration
\\  \hline
$cb\bar b\bar q$&$1/2(0^+)$& $12135-0.1i$&$37\%$&$63\%$&$0.67$&$0.83$&$0.69$&$0.82$&$0.90$&$0.41$& C.
\\
&&$12205-1i$&$27\%$&$73\%$&$0.68$&$0.94$&$0.77$&$0.98$&$1.04$&$0.46$&C.
\\
&$1/2(1^+)$&$12144-1i$&$37\%$&$63\%$&$0.68$&$0.84$&$0.69$&$0.83$&$0.91$&$0.42$&C.
\\
&$1/2(2^+)$&$12213-2i$&$37\%$&$63\%$&$0.73$&$0.91$&$0.71$&$0.89$&$0.99$&$0.40$&C.
\\ 
&&$12321-19i$&$58\%$&$42\%$&$0.61$&$0.88$&?&?&?&?&?
\\ \hline
$bc\bar c\bar q$&$1/2(0^+)$& $8875-7i$&$88\%$&$12\%$&$0.78$&$0.89$&$0.53$&$0.91$&$0.91$&$0.82$& C.
\\
&&$8950-2i$&$32\%$&$68\%$&$0.52$&$1.01$&$0.74$&$1.06$&$0.97$&$0.74$&C.
\\
&&$9004-5i$&$40\%$&$60\%$&$0.61$&$1.04$&$0.77$&$1.02$&$0.88$&$0.80$&C.
\\
&&$9030-3i$&$49\%$&$51\%$&$0.56$&$1.11$&$0.76$&$1.07$&$1.01$&$0.81$&C.
\\
&&$9078-13i$&$78\%$&$22\%$&$0.64$&$0.97$&$0.79$&$0.90$&$0.87$&$0.85$&C.
\\
&&$9096-12i$&$65\%$&$35\%$&?&?&?&?&?&?&?
\\
&$1/2(1^+)$& $8885-4i$&$74\%$&$26\%$&$0.74$&$0.85$&$0.75$&$1.04$&$1.02$&$0.94$& C.
\\ 
&&$8989-3i$&$28\%$&$72\%$&$0.49$&$1.10$&$0.80$&$1.12$&$1.02$&$0.76$&C.
\\
&&$9014-5i$&$25\%$&$75\%$&$0.55$&$0.79$&$0.63$&$0.76$&$0.93$&$0.81$&C.
\\
&&$9048-4i$&$71\%$&$29\%$&$0.50$&$1.08$&$0.69$&$1.02$&$0.99$&$0.79$&C.
\\
&&$9076-2i$&$69\%$&$31\%$&$0.55$&$1.08$&$0.64$&$1.01$&$0.97$&$0.81$&C.
\\
&&$9091-6i$&$72\%$&$28\%$&$0.57$&$1.08$&$0.68$&$0.98$&$0.95$&$0.81$&C.
\\
&&$9097-15i$&$68\%$&$32\%$&$0.58$&$0.97$&$0.74$&$0.87$&$0.88$&$0.73$&C.
\\
&&$9097-13i$&$59\%$&$41\%$&$0.58$&$0.98$&$0.68$&$0.96$&$0.98$&$0.73$&C.
\\
&$1/2(2^+)$& $9095-1i$&$64\%$&$36\%$&$0.54$&$1.11$&$0.64$&$1.06$&$1.03$&$0.83$& C.
\\ 
&&$9109-12i$&$66\%$&$34\%$&$0.75$&$1.08$&$0.91$&$1.02$&$0.91$&$0.76$&C.
\\
&&$9213-6i$&$74\%$&$26\%$&$0.88$&$1.17$&$0.53$&$1.19$&$1.13$&$0.86$&C.
\\
&&$9223-6i$&$48\%$&$52\%$&$0.81$&$1.17$&$0.85$&$1.18$&$1.10$&$0.92$&C.
\\
&&$9254-3i$&$35\%$&$65\%$&$0.70$&$1.14$&$0.89$&$1.00$&$0.95$&$0.87$&C.
\\ \hline\hline
\end{tabular}
\label{tab:QQpQpq}
\end{center}
\end{table*}

For the $cb\bar b\bar q$ system, we acquire two resonant states with quantum number $J^P=0^+$, one resonant state with $J^P=1^+$ and two resonant structures with $J^P=2^+$, respectively. Below the $D\eta_b(3S)$ threshold, the two resonant states, $T_{cbb,0}(12135)$ and $T_{cbb,0}(12205)$, exhibit compact spatial configurations, with the rms radius $r_{b\bar b}^{\rm rms}$ being the smallest among the six rms radii. The lowest state $T_{cbb,0}(12135)$ has a $67\%$ $\chi_{6_c\otimes \bar 6_c}$ color component and a nearly zero width. In the sector with spin-parity $J^P=1^+$, the compact resonant state $T_{cbb,1}(12144)$ has a similar color configuration and spatial distribution as the state $T_{cbb,0}(12135)$. Additionally, we identify two resonant states below the $D^*\Upsilon(3S)$ threshold. The compact resonant state $T_{cbb,2}(12213)$ also has a dominant $\chi_{6_c\otimes\bar 6_c}$ color component and a small width. On the other hand, the state $T_{cbb,2}(12321)$ has a width of $\Gamma=40$ MeV. Its rms radii $r_{c\bar b}^{\rm rms}$ and $r_{b\bar q}^{\rm rms}$ are similar to those of $B_c^*(2S)$ and $\bar B^*$. The remaining rms radii are relatively large but numerically unstable.

We obtain nineteen resonant states with quantum numbers $J^P=0^+$, $1^+$ and $2^+$ in $bc\bar c\bar q$ system. Among these, six resonant states with spin-parity $J^P=0^+$ lie below the $\bar B_c(3S)D$ threshold. The compact tetraquark state $T_{bcc,0}(8875)$ has the lowest mass and a dominant $\chi_{\bar 3_c\otimes 3_c}$ color component. It has a width of $\Gamma=14$~MeV and can decay into $\bar B_c D$, $\bar B\eta_c$, $\bar B_c^* D^*$, $\bar B^* J/\psi$, and $\bar B_c(2S) D$ final states. In the $bc\bar c\bar q$ system with spin-parity $J^P=1^+$, we identify eight compact resonant states within the range of $8.8$-$9.1$~GeV. For the tetraquark state $T_{bcc,1}(8885)$, the four valence quarks are distributed relatively evenly, with no significant differences in the rms radii between the quarks. In contrast, the remaining seven states exhibit a small distance between the bottom quark and anti-charm quark. Finally, we identify five compact resonant state with quantum number $J^P=2^+$ under the $\bar B_c^*(3S) D^*$ channel. The state $T_{bcc,2}(9095)$ and $T_{bcc,2}(9109)$ may be more likely to be observed experimentally and can be explored in the $\bar B_c^*D^*$, $\bar B J/\psi$, $\bar B_c^*(2S)D^*$, $\bar B^*\psi(2S)$, and $\bar B_c^* D^*(2S)$ decay channels.

\subsection{$Q Q\bar Q\bar s$}

\begin{figure*}[hbtp]
\begin{center}
\scalebox{0.6}{\includegraphics{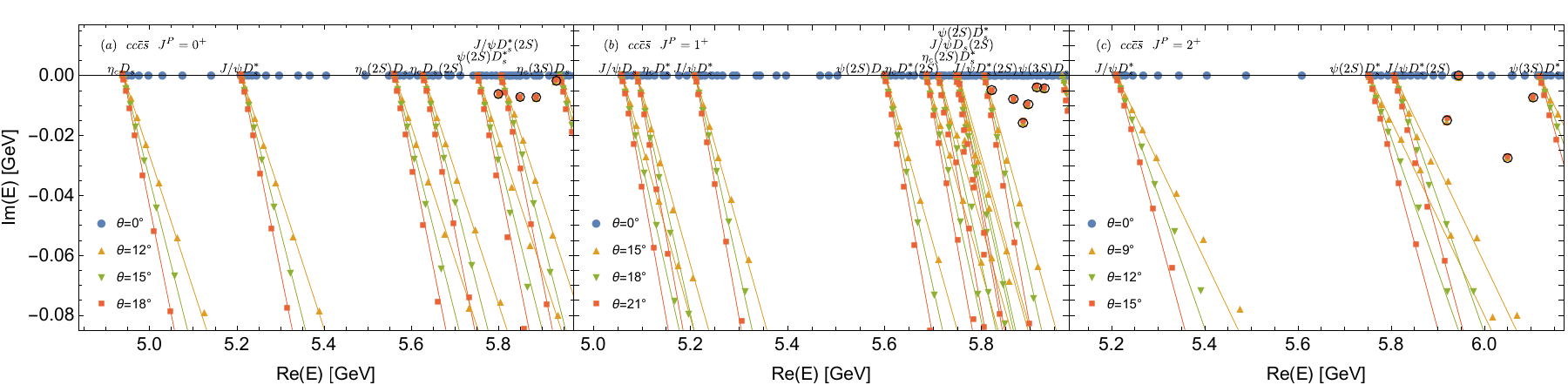}}
\end{center}
\caption{The complex eigenenergies of the $cc\bar c\bar s$ tetraquark states in the AP1 quark potential model with varying $\theta$ in the CSM.}
\label{fig:cccs}
\end{figure*}

\begin{figure*}[hbtp]
\begin{center}
\scalebox{0.6}{\includegraphics{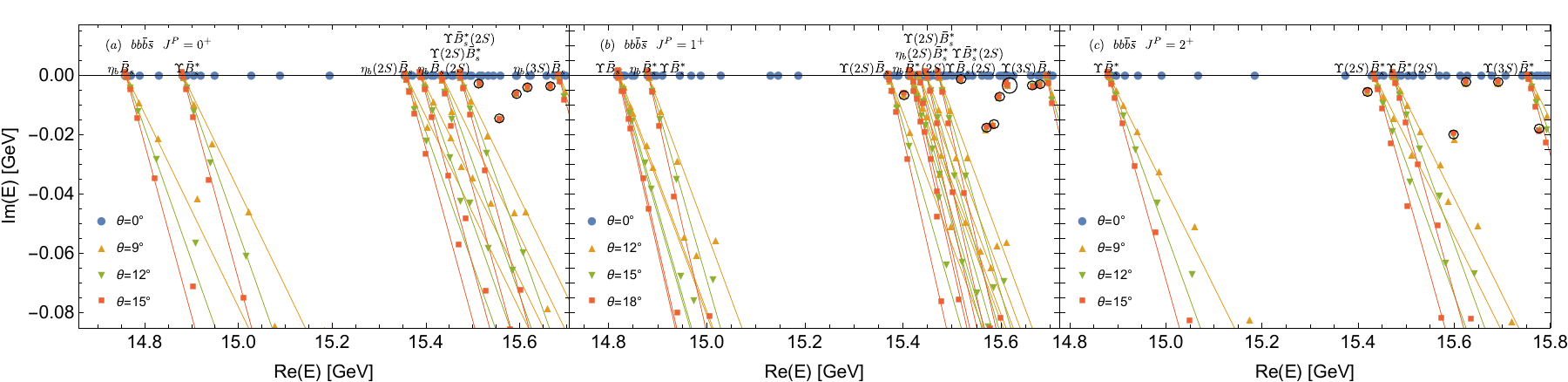}}
\end{center}
\caption{The complex eigenenergies of the $bb\bar b\bar s$ tetraquark states in the AP1 quark potential model with varying $\theta$ in the CSM.}
\label{fig:bbbs}
\end{figure*}

For the $S$-wave $Q Q\bar Q\bar s$ systems, a series of resonant states, denoted as $T_{3Q\bar s,J}(M)$, with quantum numbers $J^P=0^+$, $1^+$, and $2^+,$ are obtained by solving the complex-scaled four-body Schr\"{o}dinger equation. The distribution of complex eigenenergies for $cc\bar c\bar s$ and $bb\bar b\bar s$ systems are shown in Figs.~\ref{fig:cccs}-\ref{fig:bbbs}. The complex energies, proportions of different color configurations and rms radii of these states are summarized in Table~\ref{tab:QQQs}.

\begin{table*}[hbtp]
\begin{center}
\renewcommand{\arraystretch}{1.5}
\caption{The complex eigenenergies, proportions of two color components, and rms radii between different quarks of the $S$-wave $QQ\bar Q \bar s$ tetraquark states. The last column indicates the spatial configuration of each state.}
\begin{tabular}{c  c  c  c  c  c c  c  c c c c}
\hline\hline
system &$I(J^P)$&~~~$M-i\Gamma/2$ & $\chi_{\bar{3}_c\otimes 3_c}$~~~&~~~$\chi_{6_c\otimes \bar{6}_c}$~~~&~~~$r_{c_1\bar c}^{\rm{rms}}$~~~&~~~$r_{c_2\bar s}^{\rm{rms}}$~~~&~~~$r_{c_1 c_2}^{\rm{rms}}$~~~&~~~$r_{\bar c\bar s}^{\rm{rms}}$~~~&~~~$r_{c_1 \bar s}^{\rm{rms}}$~~~&~~~$r_{c_2\bar c}^{\rm{rms}}$~~~& Configuration
\\  \hline
$cc\bar c \bar s$&$0(0^+)$& $5801-6i$&$41\%$&$59\%$&$0.63$&$0.81$&$0.68$&$0.89$&$0.89$&$0.82$& C.
\\ 
&&$5851-8i$&$20\%$&$80\%$&$0.57$&$0.88$&$0.83$&$1.03$&$1.03$&$0.78$&C.
\\
&&$5887-7i$&$67\%$&$33\%$&$0.67$&$0.84$&$0.72$&$0.83$&$0.89$&$0.80$&C.
\\
&&$5932-2i$&$66\%$&$34\%$&$0.60$&$0.91$&$0.80$&$0.89$&$0.90$&$0.69$&C.
\\
&$0(1^+)$&$5822-5i$&$36\%$&$64\%$&$0.66$&$0.77$&$0.55$&$0.65$&$0.64$&$0.61$&C.
\\
&&$5867-8i$&$64\%$&$36\%$&$0.56$&$0.93$&$0.68$&$0.92$&$0.94$&$0.76$&C.
\\
&&$5886-15i$&$44\%$&$56\%$&$0.64$&$0.85$&$0.76$&$0.91$&$0.98$&$0.72$&C.
\\
&&$5896-10i$&$70\%$&$30\%$&$0.67$&$0.80$&$0.80$&$0.79$&$0.87$&$0.69$&C.
\\
&&$5914-4i$&$54\%$&$46\%$&$0.56$&$0.98$&$0.70$&$0.98$&$1.01$&$0.74$&C.
\\
&&$5930-4i$&$84\%$&$16\%$&$0.62$&$0.88$&$0.78$&$0.87$&$0.91$&$0.66$&C.
\\
&$0(2^+)$&$5921-14i$&$49\%$&$51\%$&$0.61$&$0.93$&$0.61$&$0.99$&$1.10$&$0.65$&C.
\\ 
&&$5945-0.3i$&$90\%$&$10\%$&$0.64$&$0.88$&$0.80$&$0.84$&$0.88$&$0.68$&C.
\\
&&$6050-27i$&$86\%$&$14\%$&$0.84$&$1.03$&$0.53$&$0.93$&$1.01$&$0.80$&C.
\\
&&$6106-7i$&$86\%$&$14\%$&$1.08$&$0.93$&$0.76$&$0.97$&$1.06$&$1.10$&C.
\\ \hline
$bb\bar b \bar s$&$0(0^+)$& $15515-3i$&$46\%$&$54\%$&$0.43$&$0.67$&$0.50$&$0.71$&$0.70$&$0.60$& C.
\\ 
&&$15558-15i$&$24\%$&$76\%$&$0.40$&$0.76$&$0.51$&$0.89$&$0.92$&$0.52$&C.
\\
&&$15594-7i$&$73\%$&$27\%$&$0.41$&$0.75$&$0.50$&$0.68$&$0.73$&$0.47$&C.
\\
&&$15618-4i$&$61\%$&$39\%$&$0.37$&$0.81$&$0.50$&$0.77$&$0.78$&$0.44$&C.
\\
&&$15667-4i$&$78\%$&$22\%$&$0.72$&$0.74$&$0.49$&$0.73$&$0.84$&$0.75$&C.
\\
&$0(1^+)$&$15402-6i$&$84\%$&$16\%$&$0.52$&$0.64$&$0.50$&$0.72$&$0.75$&$0.65$&C.
\\
&&$15519-1i$&$36\%$&$64\%$&$0.43$&$0.67$&$0.50$&$0.70$&$0.68$&$0.58$&C.
\\
&&$15571-18i$&$35\%$&$65\%$&$0.44$&$0.70$&$0.50$&$0.79$&$0.85$&$0.48$&C.
\\
&&$15584-17i$&$26\%$&$74\%$&$0.39$&$0.81$&$0.50$&$0.93$&$0.94$&$0.52$&C.
\\
&&$15596-7i$&$83\%$&$17\%$&$0.42$&$0.73$&$0.53$&$0.64$&$0.71$&$0.45$&C.
\\
&&$15612-3i$&$55\%$&$45\%$&$0.35$&$0.83$&$0.47$&$0.81$&$0.81$&$0.46$&C.
\\
&&$15615-4i$&$78\%$&$22\%$&$0.40$&$0.75$&$0.51$&$0.71$&$0.74$&$0.44$&C.
\\
&&$15665-4i$&$84\%$&$16\%$&$0.72$&$0.77$&$0.42$&$0.68$&$0.83$&$0.73$&C.
\\
&&$15680-3i$&$87\%$&$13\%$&$0.72$&$0.80$&$0.40$&$0.68$&$0.84$&$0.73$&C.
\\
&$0(2^+)$&$15421-6i$&$84\%$&$16\%$&$0.52$&$0.64$&$0.46$&$0.70$&$0.72$&$0.62$&C.
\\ 
&&$15602-21i$&$30\%$&$70\%$&$0.38$&$0.83$&$0.50$&$0.95$&$0.96$&$0.56$&C.
\\
&&$15625-3i$&$82\%$&$18\%$&$0.41$&$0.75$&$0.52$&$0.70$&$0.73$&$0.43$&C.
\\
&&$15692-3i$&$89\%$&$11\%$&$0.72$&$0.81$&$0.37$&$0.67$&$0.84$&$0.73$&C.
\\
&&$15779-18i$&$81\%$&$19\%$&$0.53$&$0.82$&$0.78$&$1.11$&$1.08$&$0.85$&C.
\\ \hline\hline
\end{tabular}
\label{tab:QQQs}
\end{center}
\end{table*}

\subsubsection{$cc\bar c \bar s$}

We obtain fourteen resonant states in the $cc\bar c \bar s$ system with quantum numbers $J^P=0^+$, $1^+$ and $2^+$. In $J^P=0^+$ system, we identify four structures below the $\eta_c(3S) D_s$ threshold within the mass interval of $5.8$-$6.0$~GeV. All these states exhibit compact spatial configurations and share five common decay channels: $\eta_cD_s$, $J/\psi D^*_s$, $\eta_c(2S)D_s$, $\eta_cD_s(2S)$, and $\psi(2S)D^*_s$. The $T_{3c\bar s,0}(5801)$ lies between the $\psi(2S)D^*_s$ and $J/\psi D_s^*(2S)$ thresholds in the complex energy plane, resembling the $T_{3c,0}(5768)$, which lies above the $\psi(2S)D^*$ threshold and below the $J/\psi D^*(2S)$ threshold. This suggests that $T_{3c\bar s,0}(5801)$ may be a strange partner state of $T_{3c,0}(5768)$. The resonant state $T_{3c\bar s,0}(5932)$ exhibits internal structures similar to those of $T_{3c,0}(5867)$. The rms radii $r_{c_1c_2}^{\rm rms}$, $r_{c_1\bar c_3}^{\rm rms}$, and $r_{c_2\bar c_3}^{\rm rms}$ between charm quarks are analogous for both states, while the rms radii between charm quarks and strange quark are smaller for $T_{3c\bar{s},0}(5932)$ compared to those of $T_{3c,0}(5867)$. The state $T_{3c\bar s,0}(5932)$ has an extra decay channel, $J/\psi D^*_s(2S)$, and its decay width is approximately $\Gamma=4$~MeV. 

We identify six resonant states with quantum number $J^P=1^+$. Their masses lie below the $\psi(3S)D_s$ threshold, ranging from $5.8$ to $6.0$~GeV. These states all have compact spatial configurations and share nine common decay channels: $J/\psi D_s$, $\eta_c D^*_s$, $J/\psi D^*_s$, $\psi(2S)D_s$, $\eta_c D^*_s(2S)$, $\eta_c(2S)D^*_s$, $J/\psi D_s(2S)$, $\psi(2S)D^*_s$, and $J/\psi D^*_s(2S)$. For the lowest state $T_{3c\bar s,1}(5822)$, a higher proportion of the color component is $\chi_{6_c\otimes \bar 6_c}$, indicating a stronger attractive interaction between diquark and antidiquark clusters, similar to the states $T_{3c,0}(5768)$ and $T_{3c\bar s,0}(5801)$. The similar rms radii between all pairs of the four quarks suggest that both the interactions within diquark(antiqiduark) clusters and the interactions between clusters are important.

For the $S$-wave $cc\bar c\bar s$ system with quantum number $J^P=2^+$, we identify four compact resonant states in complex energy plane, located on the second Riemann sheet of $J/\psi D_s^*(2S)$ channel and the first Riemann sheet of the $\psi(3S)D_s^*$ channel. These states share three same decay channels: $J/\psi D_s^*$, $\psi(2S) D_s^*$, and $J/\psi D_s^*(2S)$. Notably, the color-spin configuration $[(cc)_{\bar 3_c}^1(\bar c\bar s)_{3_c}^1]_{1c}^2$ dominates in the state $T_{3c\bar s,2}(5945)$, accounting for up to $90\%$, indicating that the spatial wave function between charm quarks is symmetric. This state is also characterized by a relatively narrow width. Furthermore, it is possible that the states $T_{3c\bar s,2}(5945)$ and $T_{3c\bar s,2}(6106)$ can be regarded as strange partners of the states $T_{3c,2}(5876)$ and $T_{3c,2}(5987)$, respectively. They all exhibit dominant $\chi_{\bar 3_c\otimes 3_c}$ color configurations, and the rms radii between the charm quarks are of similar order.

\subsubsection{$bb\bar b\bar s$}

For the $S$-wave $bb\bar b\bar s$ system, we obtain nineteen compact resonant states with quantum numbers $J^P=0^+$, $1^+$ and $2^+$. Among these, five compact resonant states possess spin-parity $J^P=0^+$. These states lie below the $\eta_b(3S) \bar B_s$ threshold and their mass spectra are concentrated in the energy range of $15.5$-$15.7$~GeV. The states $T_{3b\bar s,0}(15558)$, $T_{3b\bar s,0}(15594)$, and $T_{3b\bar s,0}(15618)$ may appear as bottom partners of $T_{3c\bar s,0}(5851)$, $T_{3c\bar s,0}(5887)$, and $T_{3c\bar s,0}(5932)$, respectively. This attribution is supported by their similar color compositions, semblable spatial distribution and the mass differences of approximately $3(m_b-m_c)$ between these pairs. These states can all decay into the following channels: $\eta_b\bar B_s$, $\Upsilon\bar B_s^*$, $\eta_b(2S)\bar B_s(2S)$, $\Upsilon(2S)\bar B_s^*$, and $\Upsilon\bar B_s^*(2S)$.
 
The lowest resonant state $T_{3b\bar s,1}(15402)$ in the $bb\bar b\bar s$ system predominantly features a $\chi_{\bar 3_c\otimes 3_c}$ color component and exhibits similar rms radii between quarks. The decay width of this state is around $12$~MeV and can be searched for in $\Upsilon\bar B_s$, $\eta_b\bar B^*_s$, $\Upsilon \bar B_s^*$, and $\Upsilon(2S)\bar B_s$ decay channels. 

For $S$-wave $bb\bar b\bar s$ system with quantum number $J^P=2^+$, five compact resonant states are obtained, all of which lie below the $\Upsilon(3S)\bar B_s^*$ threshold. Their masses range from $15.4$ to $15.8$~GeV. The states $T_{3b\bar s,2}(15421)$, $T_{3b\bar s,2}(15602)$, $T_{3b\bar s,2}(15625)$, and $T_{3b\bar s,2}(15692)$ are likely strange partners of $T_{3b,2}(15351)$, $T_{3b,2}(15544)$, $T_{3b,2}(15585)$, and $T_{3b,2}(15611)$, respectively. We note that the state $T_{3b,2}(15351)$ lies below the $\Upsilon(2S)\bar B^*$ threshold, which is consistent with $T_{3b\bar s,2}(15421)$ lying below the $\Upsilon(2S)\bar B_s^*$  threshold. In particular, the state $T_{3b\bar s,2}(15602)$ has a relatively large width of about $42$~MeV. It can decay into the $\Upsilon \bar B_s^*$, $\Upsilon(2S)\bar B_s^*$, and $\Upsilon \bar B_s^*(2S)$ channels.

\subsection{$QQ\bar Q^\prime\bar s$}

\begin{figure*}[hbtp]
\begin{center}
\scalebox{0.6}{\includegraphics{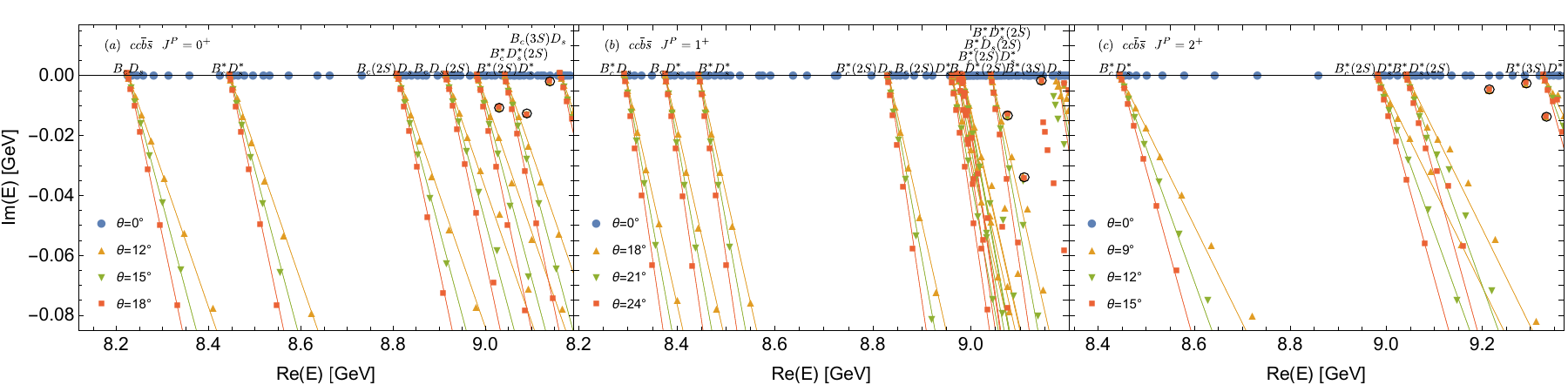}}
\end{center}
\caption{The complex eigenenergies of the $cc\bar b\bar s$ tetraquark states in the AP1 quark potential model with varying $\theta$ in the CSM.}
\label{fig:ccbs}
\end{figure*}

\begin{figure*}[hbtp]
\begin{center}
\scalebox{0.6}{\includegraphics{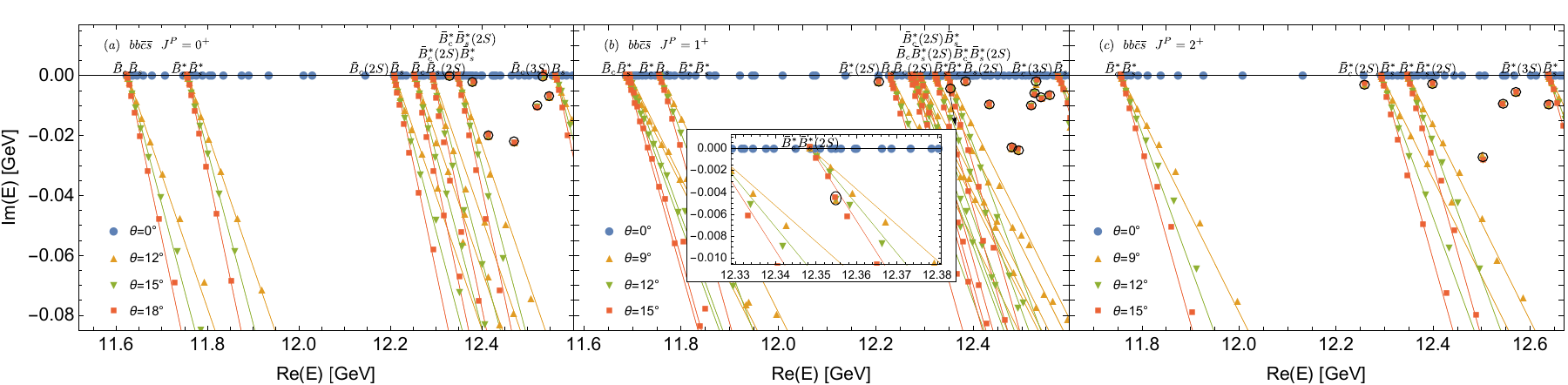}}
\end{center}
\caption{The complex eigenenergies of the $bb\bar c\bar s$ tetraquark states in the AP1 quark potential model with varying $\theta$ in the CSM.}
\label{fig:bbcs}
\end{figure*}

The complex eigenenergies of the $S$-wave $QQ\bar Q^\prime\bar s$ systems are shown in Figs.~\ref{fig:ccbs}-\ref{fig:bbcs}, respectively. We obtain a series of resonant states, labeled as $T_{QQ Q^\prime\bar s,J}(M)$, in the following discussions. For these states, the complex eigenenergies, proportions of two color components and rms radii of different quarks are presented in Table~\ref{tab:QQQps}.

\begin{table*}[hbtp]
\begin{center}
\renewcommand{\arraystretch}{1.5}
\caption{The complex eigenenergies, proportions of two color components, and rms radii between different quarks of the $S$-wave $QQ\bar Q^\prime \bar s$ tetraquark states. The last column indicates the spatial configuration of each state.}
\begin{tabular}{c c  c  c  c  c c  c  c c c c}
\hline\hline
system &$I(J^P)$&~~~$M-i\Gamma/2$ & $\chi_{\bar{3}_c\otimes 3_c}$~~~&~~~$\chi_{6_c\otimes \bar{6}_c}$~~~&~~~$r_{Q_1\bar Q_3}^{\rm{rms}}$~~~&~~~$r_{Q_2\bar q}^{\rm{rms}}$~~~&~~~$r_{Q_1 Q_2}^{\rm{rms}}$~~~&~~~ $r_{\bar Q_3\bar q}^{\rm{rms}}$~~~&~~~$r_{Q_1 \bar q}^{\rm{rms}}$~~~&~~~$r_{Q_2\bar Q_3}^{\rm{rms}}$~~~& Configuration
\\  \hline
$cc\bar b\bar s$&$0(0^+)$& $9030-11i$&$37\%$&$63\%$&$0.47$&$0.88$&$0.65$&$0.92$&$1.00$&$0.67$& C.
\\
&&$9091-13i$&$54\%$&$46\%$&$0.51$&$0.91$&$0.76$&$0.69$&$0.82$&$0.81$&C.
\\
&&$9140-2i$&$24\%$&$76\%$&$0.55$&$0.82$&$0.76$&$0.79$&$0.82$&$0.59$&C.
\\
&$0(1^+)$&$9075-13i$&$57\%$&$43\%$&$0.43$&$0.99$&?&?&?&?&?
\\
&&$9109-35i$&$16\%$&$84\%$&$0.51$&$0.84$&$0.58$&$1.12$&$1.22$&$0.34$&C.
\\
&&$9145-2i$&$15\%$&$85\%$&$0.55$&$0.83$&$0.77$&$0.82$&$0.85$&$0.60$&C.
\\
&$0(2^+)$&$9215-5i$&$97\%$&$3\%$&$0.61$&$0.83$&$0.81$&$0.69$&$0.84$&$0.62$&C.
\\ 
&&$9292-3i$&$89\%$&$11\%$&$0.87$&$0.99$&$0.62$&$0.74$&$1.03$&$0.90$&C.
\\
&&$9334-14i$&$83\%$&$17\%$&$0.64$&$0.96$&$0.64$&$0.85$&$0.91$&$0.68$&C.
\\ \hline
$bb\bar c\bar s$&$0(0^+)$& $12330-0.3i$&$53\%$&$47\%$&$0.57$&$0.67$&$0.52$&$0.82$&$0.72$&$0.67$& C.
\\
&&$12382-2i$&$76\%$&$24\%$&$0.55$&$0.69$&$0.53$&$0.76$&$0.72$&$0.58$&C.
\\
&&$12415-20i$&$57\%$&$43\%$&$0.48$&$0.82$&$0.52$&$0.82$&$0.84$&$0.49$&C.
\\
&&$12473-23i$&$44\%$&$56\%$&$0.59$&$0.89$&$0.38$&$0.97$&$0.93$&$0.62$&C.
\\
&&$12522-11i$&$69\%$&$31\%$&$0.81$&$0.94$&$0.34$&$0.80$&$0.94$&$0.82$&C.
\\
&&$12537$&$29\%$&$71\%$&$0.58$&$0.82$&$0.77$&$0.96$&$0.87$&$0.73$&C.
\\
&&$12549-7i$&$77\%$&$23\%$&$0.87$&$0.72$&$0.50$&$1.02$&$0.81$&$0.87$&C.
\\
&$0(1^+)$& $12207-2i$&$90\%$&$10\%$&$0.69$&$0.69$&$0.40$&$0.71$&$0.74$&$0.73$& C.
\\ 
&&$12355-4i$&$65\%$&$35\%$&$0.54$&$0.72$&?&?&?&?&?
\\
&&$12385-2i$&$76\%$&$24\%$&$0.55$&$0.69$&$0.55$&$0.74$&$0.71$&$0.56$&C.
\\
&&$12434-10i$&$77\%$&$23\%$&$0.53$&$0.81$&$0.48$&$0.83$&$0.82$&$0.61$&C.
\\
&&$12481-24i$&$40\%$&$60\%$&$0.57$&$0.90$&$0.39$&$0.98$&$0.94$&$0.61$&C.
\\
&&$12494-25i$&$59\%$&$41\%$&$0.79$&$0.89$&$0.36$&$0.84$&$0.84$&$0.86$&C.
\\
&&$12521-10i$&$43\%$&$57\%$&$0.84$&$0.93$&$0.31$&$0.77$&$0.97$&$0.82$&C.
\\
&&$12528-6i$&$76\%$&$24\%$&$0.85$&$0.93$&$0.42$&$0.84$&$0.95$&$0.85$&C.
\\
&&$12532-2i$&$23\%$&$77\%$&$0.53$&$0.81$&$0.78$&$0.93$&$0.85$&$0.66$&C.
\\
&&$12542-8i$&$62\%$&$38\%$&$0.68$&$0.79$&$0.53$&$0.96$&$0.83$&$0.69$&C.
\\
&&$12559-7i$&$82\%$&$18\%$&$0.86$&$0.77$&$0.45$&$0.99$&$0.81$&$0.86$&C.
\\
&$0(2^+)$& $12260-3i$&$91\%$&$9\%$&$0.71$&$0.70$&$0.40$&$0.76$&$0.75$&$0.75$& C.
\\ 
&&$12400-3i$&$78\%$&$22\%$&$0.56$&$0.69$&$0.55$&$0.72$&$0.69$&$0.57$&C.
\\
&&$12504-28i$&$34\%$&$66\%$&$0.58$&$0.88$&$0.40$&$0.99$&$0.94$&$0.63$&C.
\\
&&$12545-10i$&$57\%$&$43\%$&$0.78$&$0.94$&$0.35$&$0.83$&$0.94$&$0.79$&C.
\\
&&$12572-6i$&$87\%$&$13\%$&$0.88$&$0.81$&$0.41$&$0.96$&$0.83$&$0.87$&C.
\\
&&$12640-9i$&$75\%$&$25\%$&$0.68$&$0.80$&$0.80$&$0.80$&$0.78$&$0.71$&C.
\\ \hline\hline
\end{tabular}
\label{tab:QQQps}
\end{center}
\end{table*}

We acquire thirty-three resonant states in $S$-wave $QQ\bar Q^\prime\bar s$ systems below the $M_1(3S)M_2(1S)$ threshold. For the $cc\bar b\bar s$ tetraquark system, we obtain three resonant states with quantum number $J^P=0^+$. The lowest compact state $T_{ccb\bar s,0}(9030)$, with a width of $\Gamma=22$~MeV, lies below the $B_c^*D_s^*(2S)$ threshold, and serves as the strange analogue of the $T_{ccb,0}(9015)$, which is located below the $B_c^*D^*(2S)$ threshold. This state contains approximately $60\%$ of the $\chi_{6_c\otimes \bar 6_c}$ color component, consistent with the $T_{ccb,0}(9015)$. It can decay into the $B_c D_s$, $B_c^* D_s^*$, $B_c(2S)D_s$, $B_cD_s(2S)$ and $B_c^*(2S)D_s^*$ channels. Another higher compact state, $T_{ccb\bar s,0}(9140)$, is also the strange partner of $T_{ccb,0}(9057)$. The mass difference between these two states is around $80$ MeV. For the states with quantum number $J^P=1^+$, two resonant states, $T_{ccb\bar s,1}(9109)$ and $T_{ccb\bar s,1}(9145)$, exhibit compact spatial configurations. The resonant state $T_{ccb\bar s,1}(9075)$ is close to the continuum states located on the $2\theta$-lines starting from the $B_c^*D_s(2S)$ threshold. As a result, the rms radii $r_{c_1\bar b}^{\rm rms}$ and $r_{c_2\bar s}^{\rm rms}$ are similar to the spatial sizes of dimeson $B_c^*D_s(2S)$, while the other rms radii are numerically unstable and vary dramatically with $\theta$. The $cc\bar b\bar s$ system with quantum number $J^P=2^+$ includes three compact tetraquark structures: $T_{ccb\bar s,2}(9215)$, $T_{ccb\bar s,2}(9292)$, and $T_{ccb\bar s,2}(9334)$. These states all feature the $[(cc)_{\bar 3_c}^1(\bar b\bar s)_{3_c}^1]_{1_c}^2$ dominant color-spin configurations and are likely strange analogues of the states $T_{ccb,2}(9150)$, $T_{ccb,2}(9201)$ and  $T_{ccb,2}(9254)$.

For $S$-wave $bb\bar c\bar s$ system, we identify more resonant states, including seven compact tetraquark states with quantum number $J^P=0^+$, eleven resonant states with $J^P=1^+$, and six compact resonant states with $J^P=2^+$. Similar to $T_{bbc,0}(12248)$, the $T_{bbc\bar s,0}(12330)$ has a very small width. In the sector with quantum number $J^P=1^+$, we note that the compact tetraquark state $T_{bbc\bar s,1}(12207)$, with a width of $\Gamma=4$ MeV, may be the strange partner state of the $T_{bbc,1}(12124)$. Both states predominantly exhibit the $\chi_{\bar 3_c\otimes 3_c}$ color component, and the spatial sizes of the heavy quarks are semblable. Below the mass of $T_{bbc\bar s,1}(12207)$, there are three dimeson decay channels: $\bar B_c\bar B_s^*$, $\bar B_c^*\bar B_s$, and $\bar B_c^*\bar B_s^*$. The four compact resonant states with higher spin $J^P=2^+$ are denoted as $T_{bbc\bar s,2}(12260)$, $T_{bbc\bar s,2}(12545)$, $T_{bbc\bar s,2}(12572)$, and $T_{bbc\bar s,2}(12640)$. These states may serve as the strange analogues of the $T_{bbc,2}(12180)$, $T_{bbc,2}(12483)$, $T_{bbc,2}(12532)$, and $T_{bbc,2}(12579)$. The state $T_{bbc\bar s,2}(12260)$ has a symmetric spatial wave function between the bottom  quarks, similar to the $T_{ccb,2}(12180)$, and decays exclusively via the $\bar B_c^*\bar B_s^*$ channel. In contrast, the states $T_{bbc\bar s,2}(12545)$, $T_{bbc\bar s,2}(12572)$, and $T_{bbc\bar s,2}(12640)$ have additional decay channels, including $\bar B_c^*(2S)\bar B_s^*$ and $\bar B_c^*\bar B_s^*(2S)$.

\subsection{$QQ^\prime\bar Q^\prime\bar s$}

\begin{figure*}[hbtp]
\begin{center}
\scalebox{0.6}{\includegraphics{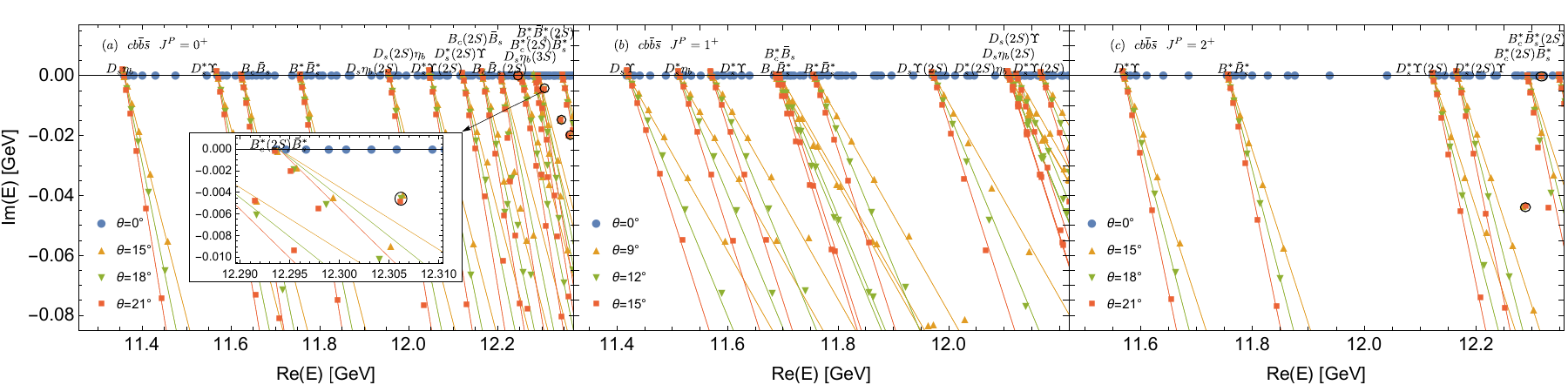}}
\end{center}
\caption{The complex eigenenergies of the $cb\bar b\bar s$ tetraquark states in the AP1 quark potential model with varying $\theta$ in the CSM.}
\label{fig:cbbs}
\end{figure*}

\begin{figure*}[hbtp]
\begin{center}
\scalebox{0.6}{\includegraphics{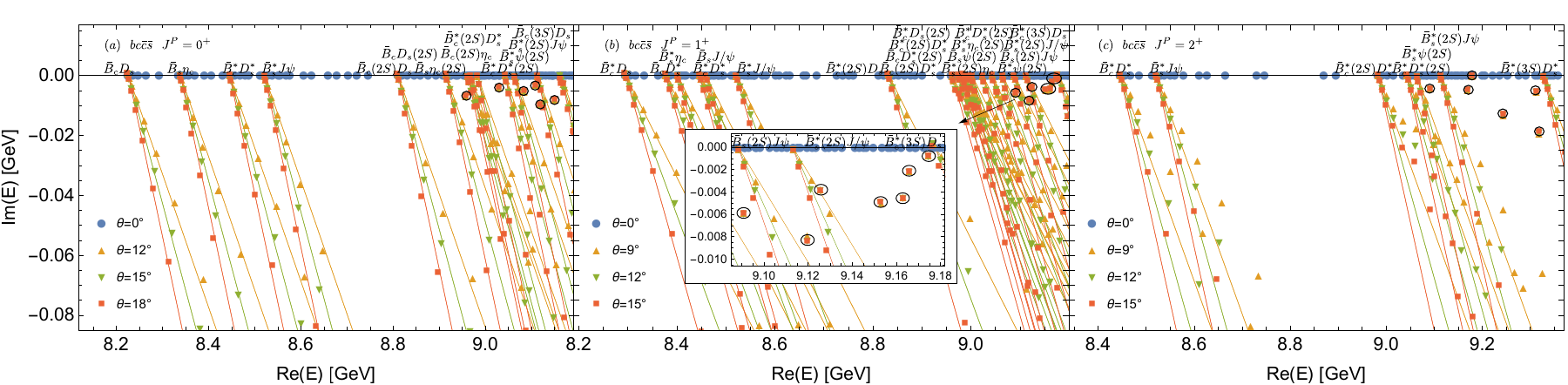}}
\end{center}
\caption{The complex eigenenergies of the $bc\bar c\bar c$ tetraquark states in the AP1 quark potential model with varying $\theta$ in the CSM.}
\label{fig:bccs}
\end{figure*}

The complex eigenenergies of the $S$-wave $QQ^\prime\bar Q^\prime\bar s$ systems are shown in Figs.~\ref{fig:cbbs}-\ref{fig:bccs}, respectively. We obtain a series of resonant states, labeled as $T_{QQ^\prime Q^\prime\bar s,J}(M)$, which will be discussed in the following sections. The complex eigenenergies, proportions of two color components, and rms radii of different quarks are listed in Table~\ref{tab:QQpQps}.

\begin{table*}[hbtp]
\begin{center}
\renewcommand{\arraystretch}{1.5}
\caption{The complex eigenenergies, proportions of two color components, and rms radii between different quarks of the $S$-wave $QQ^\prime \bar Q^\prime \bar s$ tetraquark states. The last column indicates the spatial configuration of each state.}
\begin{tabular}{c c  c  c  c  c c  c  c c c c}
\hline\hline
system &$I(J^P)$&~~~$M-i\Gamma/2$~~~&~~~$\chi_{\bar{3}_c\otimes 3_c}$~~~&~~~$\chi_{6_c\otimes \bar{6}_c}$~~~&~~~$r_{Q_1\bar Q_3}^{\rm{rms}}$~~~&~~~$r_{Q_2\bar q}^{\rm{rms}}$~~~&~~~$r_{Q_1 Q_2}^{\rm{rms}}$~~~&~~~ $r_{\bar Q_3\bar q}^{\rm{rms}}$~~~&~~~$r_{Q_1 \bar q}^{\rm{rms}}$~~~&~~~$r_{Q_2\bar Q_3}^{\rm{rms}}$~~~& Configuration
\\  \hline
$cb\bar b\bar s$&$0(0^+)$& $12247-1i$&$33\%$&$67\%$&$0.61$&$0.69$&$0.65$&$0.70$&$0.76$&$0.42$& C.
\\
&&$12306-5i$&$39\%$&$61\%$&$0.67$&$0.69$&$0.61$&$0.66$&$0.77$&$0.42$&C.
\\
&&$12344-15i$&$49\%$&$51\%$&$0.55$&$0.71$&$0.66$&$0.76$&$0.89$&$0.51$&C.
\\
&&$12362-20i$&$42\%$&$58\%$&$0.52$&$0.81$&?&?&?&?&?
\\
&$0(2^+)$&$12286-44i$&$87\%$&$13\%$&$0.82$&$0.91$&$0.67$&$0.74$&$0.72$&$0.61$&C.
\\ 
&&$12317-0.2i$&$33\%$&$67\%$&$0.67$&$0.72$&$0.67$&$0.73$&$0.82$&$0.45$&C.
\\ \hline
$bc\bar c\bar s$&$0(0^+)$& $8959-7i$&$83\%$&$17\%$&$0.74$&$0.77$&$0.56$&$0.79$&$0.75$&$0.77$& C.
\\
&&$9030-4i$&$28\%$&$72\%$&$0.51$&$0.87$&$0.70$&$0.90$&$0.80$&$0.69$&C.
\\
&&$9084-5i$&$32\%$&$68\%$&$0.39$&$0.85$&$0.63$&$0.88$&$0.85$&$0.72$&C.
\\
&&$9109-4i$&$50\%$&$50\%$&$0.57$&$0.82$&$0.57$&$0.83$&$0.79$&$0.73$&C.
\\
&&$9120-10i$&$80\%$&$20\%$&?&?&?&?&$0.66$&$0.72$&?
\\
&&$9151-8i$&$75\%$&$25\%$&$0.54$&$0.87$&$0.71$&$0.75$&$0.68$&$0.76$&C.
\\
&$0(1^+)$& $9091-6i$&$23\%$&$77\%$&$0.37$&$0.87$&$0.70$&$0.95$&$0.92$&$0.79$& C.
\\ 
&&$9120-8i$&$57\%$&$43\%$&?&?&?&?&$0.69$&$0.69$&?
\\
&&$9126-4i$&$70\%$&$30\%$&$0.61$&$0.90$&$0.69$&$0.84$&$0.76$&$0.76$&C.
\\
&&$9153-5i$&$78\%$&$22\%$&$0.54$&$0.87$&$0.67$&$0.74$&$0.67$&$0.75$&C.
\\
&&$9163-5i$&$83\%$&$17\%$&$0.59$&$0.89$&$0.70$&$0.77$&$0.65$&$0.72$&C.
\\
&&$9166-2i$&$72\%$&$28\%$&$0.58$&$0.80$&$0.59$&$0.82$&$0.81$&$0.78$&C.
\\
&&$9175-1i$&$72\%$&$28\%$&$0.58$&$0.82$&$0.61$&$0.86$&$0.83$&$0.81$&C.
\\
&$0(2^+)$& $9092-5i$&$18\%$&$82\%$&?&?&?&?&$0.88$&$0.71$& ?
\\ 
&&$9171-5i$&$82\%$&$18\%$&$0.57$&$0.88$&$0.70$&$0.75$&$0.67$&$0.73$&C.
\\
&&$9180-0.2i$&$67\%$&$33\%$&$0.58$&$0.82$&$0.57$&$0.83$&$0.81$&$0.78$&C.
\\
&&$9244-13i$&$55\%$&$45\%$&$0.58$&$1.01$&$0.56$&$1.01$&$0.90$&$0.59$&C.
\\
&&$9313-5i$&$85\%$&$15\%$&$1.08$&$0.87$&$0.68$&$0.97$&$0.96$&$1.07$&C.
\\
&&$9318-19i$&$44\%$&$56\%$&$0.36$&$0.99$&$0.68$&$1.02$&$0.94$&$0.69$&C.
\\ \hline\hline
\end{tabular}
\label{tab:QQpQps}
\end{center}
\end{table*}

For the $cb\bar b\bar s$ system, we obtain four resonant states with spin-parity $J^P=0^+$ and two resonant states with $J^P=2^+$. The compact resonant state $T_{cbb\bar s,0}(12135)$ exhibits a $63\%$  $\chi_{6_c\otimes \bar 6_c}$ color configuration and a very small width, similar to that of the state $T_{cbb,0}(12247)$. The state $T_{cbb\bar s,2}(12317)$ is a compact tetraquark state with a large width of $\Gamma=88$ MeV. It can decay into $D_s^*\Upsilon$, $B_c^*\bar B_s^*$, $D_s^*\Upsilon(2S)$ and $D_s^*(2S)\Upsilon$ channels. For the states with quantum number $J^P=1^+$, we have only computed the complex eigenenergies of scattering states below the $D_s^*\Upsilon(2S)$ threshold. In the higher energy region, the distribution of the complex eigenenergies from the complex-scaled Schr\"{o}dinger equation at different scaling angles is quite disordered due to the limitation of numerical precision. 

For the $bc\bar c\bar s$ system, we obtain six resonant states with $J^P=0^+$, seven resonant states with $J^P=1^+$, and six resonant states with $J^P=2^+$, all below the $M_1(3S)M_2(1S)$ threshold. The lowest state $T_{bcc\bar s,0}(8959)$ is a compact tetraquark state. Its spatial distribution is characterized by the smallest rms radius between the bottom and the charm quarks, while the other rms radii are relatively close. This is somewhat similar to that of the state $T_{bcc,0}(8875)$. This state has a width of around $\Gamma=14$~MeV and can decay into $\bar B_c D_s$, $\bar B_s\eta_c$, $\bar B_c^* D_s^*$, $\bar B_s^* J/\psi$, $\bar B_c(2S)D_s$, and $\bar B_cD_s(2S)$ channels. The compact resonant state $T_{bcc\bar s,0}(9151)$ is the strange partner of $T_{bcc,0}(9078)$, with a mass difference of approximately $70$~MeV. The compact resonant state $T_{bcc\bar s,1}(9091)$ with a dominant $\chi_{6_c\otimes \bar 6_c}$ color component, has a width of about $\Gamma=12$~MeV and can be observed in experiments through the $\bar B_s^*\psi(2S)$, $\bar B_c^*D_s^*(2S)$, and $\bar B_s^*\eta_c(2S)$ channels, among others. The resonant state $T_{bcc\bar s,2}(9092)$ is sandwiched between the $\bar B_s^*\psi(2S)$ and $\bar B_s^*(2S)J/\psi$ channels. The rms radii $r_{b\bar s}^{\rm rms}$ and $r_{c\bar c}^{\rm rms}$ for this state are larger than the radii of $\bar B_s^*$ and $J/\psi$, while smaller than those of $B_s^*(2S)$ and $\psi(2S)$. Additionally, the other radii between quarks have considerable instability.

\section{summary}
\label{sec:summary}
In this work, we investigate the mass spectrum of $S$-wave triply heavy tetraquark systems with quantum numbers $J^P=0^+$, $1^+$, and $2^+$ using the AP1 quark potential model. We consider a complete set of color-spin basis functions and three different sets of Jacobi coordinates, with fourteen Gaussian basis functions for each coordinate. The complex-scaled four-body Schr\"{o}dinger equation is solved employing the complex scaling method and the Gaussian expansion method, allowing us to identify possible exotic states.

We obtain a series of resonant states in $cc\bar c\bar q/\bar s$, $bb\bar b\bar q/\bar s$, $cc\bar b\bar q/\bar s$, $bb\bar c\bar q/\bar s$, $cb\bar b\bar q/\bar s$ and $bc\bar c\bar q/\bar s$ systems, but no bound state. The mass spectrum of these states ranges from $5.6$ to $15.8$~GeV. Among them, certain resonant states, such as $T_{bbc,0}(12248)$, $T_{cbb,0}(12135)$, $T_{3c\bar s,2}(5945)$, $T_{bbc\bar s,0}(12330)$,
 $T_{bbc\bar s,0}(12537)$, $T_{cbb\bar s,2}(12317)$, and $T_{bcc\bar s,2}(9180)$, have widths less than $1$~MeV. There are several possible reasons for the narrow width: (a) The overlap between the initial and final spatial wave functions could be very small, leading to an extremely narrow width; (b) The color mixing elements via color-electric terms between the initial and final states might be suppressed; (c) The potential couplings to $P$-wave dimeson channels are suppressed, as $P$-wave channels are not explicitly considered in our calculation. The exact reasons require further analysis.

In order to describe the spatial configurations of these resonant states, we use two sets of rms radii as indicators to distinguish between meson molecular states and the compact tetraquark states. For the $S$-wave $cc\bar c\bar q/\bar s$, $bb\bar b\bar q/\bar s$, $cc\bar b\bar q/\bar s$, and $bb\bar c\bar q/\bar s$ systems, which possess a pair of identical particles, we adopt the newly defined rms radius $r_{ij}^{\rm rms,N}$, based on the decomposed wave function. This eliminates the ambiguity arising from the antisymmetrization of identical particles and allows for a better identification of the molecular configuration. For $cb\bar b\bar q/\bar s$ and $bc\bar c\bar q/\bar s$ systems, which do not contain identical particles, we employ the conventional rms radius $r_{ij}^{\rm rms,C}$ to depict the spatial distribution of resonant states. Among these states, we identify a molecular candidate $T_{3c,2}(5819)$. Its rms radii $r_{c_1\bar c}^{\rm rms}$ and $r_{c_2\bar q}^{\rm rms}$ are slightly larger than rms radii of $J/\psi $ and $D^*$, but smaller than those of $\psi(2S)$ and $D^*(2S)$. Meanwhile, these radii are much smaller than the other radii, suggesting that this state may contain a molecular component. Additionally, several resonant states, such as $T_{3c,0}(5809)$, $T_{3b,0}(15511)$, $T_{ccb,2}(8909)$, $T_{cbb,2}(12321)$, $T_{bcc,0}(9096)$, $T_{ccb\bar s,1}(9075)$, $T_{bbc\bar s,1}(12355)$, $T_{cbb\bar s,0}(12362)$, $T_{bcc\bar s,0}(9120)$, $T_{bcc\bar s,1}(9120)$, and $T_{bcc\bar s,2}(9092)$,  exhibit strong coupling to the continuum states located on the $2\theta$-lines, which result in numerical instability in their rms radii. The remaining resonant states show compact spatial configurations. For most of these states, the rms radii between heavy quarks are smaller than those between the heavy and light quarks, which resembles the spatial distribution of the tritium atom in QED.

For $S$-wave $QQ^{(\prime)}\bar Q^{(\prime)}\bar q/\bar s$ systems, the resonant state $T_{3c,1}(5634)$ is most likely to be observed in experiments due to its lowest-lying mass and narrow width, which would produce a prominent signal peak. We suggest searching for this state in the $J/\psi D$, $\eta_c D^*$ and $J/\psi D^*$ decay channels. Furthermore, experimental investigations of the $T_{3c,1}(5634)$ state would not only provide valuable experimental data, but also offer an opportunity to verify the accuracy of our predictions and enhance our understanding of QCD.

%=====================================================================================
%=====================================================================================
%=====================================================================================
\section*{ACKNOWLEDGMENTS}
%=====================================================================================
%=====================================================================================
%=====================================================================================

This project is supported by the China Postdoctoral Science Foundation under Grants No.~2024M750049, and
the National Natural Science Foundation of China under Grants No.~12475137. The computational resources were supported by High-performance Computing Platform of Peking University.

%%%%%%%%%%%%%%%%%%%%%%%%%%%%%%%%%%%%%%%%%%%%%%%%%%%%%%%%%%%%%%%%%%%%%%%%%%%%%%

%%%%%%%%%%%%%%%%%%%%%%%%%%%%%%%%%%%%%%%%%%%%%%%%%%%%%%%%%%%%%%%%%%%%%%%%%%%%%%%
%%

\end{document}